\pdfoutput=1
\documentclass{article}
\usepackage{caption}
\usepackage{subcaption}
\usepackage{graphicx}
\usepackage{amsmath}
\usepackage{geometry}
\usepackage{authblk}
\usepackage{hyperref}
\usepackage[T1]{fontenc}
\usepackage[utf8]{inputenc}
\usepackage{float}
\usepackage{csquotes}
\usepackage[version=4]{mhchem}
\usepackage{tikz}
\usepackage{amssymb}
\usetikzlibrary{backgrounds,fit,decorations.pathreplacing}
\usepackage{authblk}
\usepackage{multibib}
\newcites{latex}{Supplementary References}
\usepackage{tabularx}

\title{
Towards Perturbation Theory Methods on a Quantum Computer
}
\author[1,2]{Junxu Li}
\author[3]{Barbara Jones}
\author{Sabre Kais\thanks{Email: kais@purdue.edu}}
\affil{Department of Chemistry, Department of Physics and Astronomy, and

Purdue Quantum Science and Engineering Institute

Purdue University, West Lafayette, IN 47907, United States}
\affil[2]{
Department of Physics,
College of Sciences,

Northeastern University,
Shenyang 110819, China
}
\affil[3]{
IBM Quantum, San Jose, CA 95120, USA
}
\begin{document}
\maketitle
	
\begin{abstract}
%Perturbation theory (PT) might be one of the most powerful and fruitful tools for both physicists and chemists, which has caused an explosion of applications, including the blooming of atomic and subatomic physics.
Perturbation theory (PT), used in a wide range of fields, is a powerful tool for  approximate solutions to complex problems, starting from the exact solution of a related, simpler problem. Advances in quantum computing, especially over the last several years, provide 
opportunities for alternatives to classical methods.  
Here we present a general quantum circuit estimating both the energy and eigenstates corrections that is far superior to the classical version when estimating second order energy corrections.
%Here we present a general quantum circuit that estimates both the energy and eigenstates corrections far superior than the classical version while estimating the second-order energy corrections.
%We demonstrate our approach on the two-site extended Hubbard model. 
%Results obtained from IBM's quantum hardware are presented, in addition to the numerical simulations based on Qiskit simulators.
We demonstrate our approach as applied to the two-site extended Hubbard model.
In addition to numerical simulations based on qiskit, results on IBM's quantum hardware are also presented.
%Unlike the popular quantum variational circuit,
%Unlike the popular heuristic quantum algorithms,
%there is no training or optimizing process in our circuit, and all parameters are derived from the unperturbed Hamiltonian.
Our work offers a general approach to study complex systems with quantum devices, with no training or optimization process needed to obtain the perturbative terms, which can be generalized to other Hamiltonian systems both in chemistry and physics.
\end{abstract}

\section*{Introduction}
\label{introduction}
% A brief review of the perturbation theory in quantum mechanics
Historically, Schr\"{o}dinger's techniques presented in 1926\cite{schrodinger1926undulatory} represent the first important application of perturbation theory (PT) for quantum systems, to obtain quantum eigenenergies. With the expansion of theory for atomic and subatomic physics in the first half of the 20th century, PT methods led to a wide variety of applications, such as hyperfine structure\cite{cowan1981theory} and the Zeeman \cite{condon1935theory} and Stark effects\cite{epstein1926stark}.
Paul Dirac, studying the emission and absorption of radiation in 1927\cite{dirac1927quantum}, developed a PT result {that} became Fermi's golden rule.
 In quantum field theory, Richard Feynman introduced the diagrams known by his name\cite{feynman2018theory}, which represent the perturbative contributions to transition amplitudes. 
PT is in addition a powerful tool for chemists\cite{herschbach2012dimensional, germann1993large, kais19941, kais2003finite}.
A typical example is M\o ller–Plesset perturbation theory (MP)\cite{moller1934note}, where the difference between the exact Hamiltonian and the Hartree–Fock is included as a perturbation.

% A brief review of the success and advancement of quantum computing
We now turn to a brief review of recent advancements in quantum computing.
%Let's now briefly review recent advancements in quantum computing.
In 2019, Google claimed quantum supremacy with their programmable superconducting processor, progressing on the path to full-scale quantum computing\cite{arute2019quantum}.
In 2020, the quantum computational advantage was once again claimed on a photonic quantum computer\cite{zhong2020quantum}.
The fast-paced progress of hardware has resulted in a significant increase in quantum simulation\cite{bacon2001universal, rost2020simulation, eddins2022doubling, cohn2020minimal} and error mitigation\cite{kandala2019error} on noisy intermediate-scale quantum (NISQ)  devices\cite{bharti2022noisy}.
State-of-art variational quantum circuits also attract great interest due to their efficiency and flexibility, leading to a variety of applications ranging from data classification\cite{sajjan2021quantum, cong2019quantum, li2021quantum, amin2018quantum} to electronic structure calculations\cite{carleo2017solving, xia2020qubit, xia2018quantum}.
This progress holds the potential for PT methods to be used as an application on quantum devices.
% This progress creates promise for PT methods as an application on quantum devices. 

% Outline of this article
In this paper, we propose a universal quantum circuit implementation for time-independent PT, or as often termed, Rayleigh–Schr\"odinger PT.
Consider the Hamiltonian
\begin{equation}
    H = H_0 + \lambda V
\end{equation}
where $H_0$ is the original Hamiltonian, $V$ represents the perturbation and $\lambda\ll 1$.
Denoting the eigenstates and corresponding energy levels of $H_0$ as $|\psi_n^{(0)}\rangle$ and $E_n^{(0)}$, we have $H_0|\psi_n^{(0)}\rangle = E_n^{(0)}|\psi_n^{(0)}\rangle$.
Using time-independent PT leads to the following approximation\cite{griffiths2018introduction}
\begin{equation*}
    E_n = E_n^{(0)} + \lambda E_n^{(1)} + \lambda^2 E_n^{(2)} + O(\lambda^3)
\end{equation*}
\begin{equation*}
    |\psi_n\rangle = |\psi_n^{(0)}\rangle + \lambda |\psi_n^{(1)}\rangle + O(\lambda^2)
\end{equation*}
where the first order correction of eigenstates $|\psi_n^{(1)}\rangle$ and the first and second order corrections of energy $\lambda E_n^{(1,2)}$ are included.
Mathematically, we have the first-order correction as
\begin{equation}
    E_n^{(1)}
    =
    \langle \psi_n^{(0)}|V|\psi_n^{(0)}\rangle
    \label{eq_energy1}
\end{equation}
\begin{equation}
    |\psi_n^{(1)}\rangle
    =
    \sum_{m\neq n}
    {
    \frac{\langle \psi_m^{(0)}|V|\psi_n^{(0)}\rangle}{E_n^{(0)}-E_m^{(0)}}
    |\psi_m^{(0)}\rangle
    }
    \label{eq_state1}
\end{equation}
and the second-order correction is 
\begin{equation}
    E_n^{(2)}
    =
    \sum_{m\neq n}
    {
    \frac{\left|\langle \psi_m^{(0)}|V|\psi_n^{(0)}\rangle\right|^2}
    {E_n^{(0)}-E_m^{(0)}}
    \label{eq_energy2}
    }
\end{equation}
In our approach, simple measurements can be used to estimate the corrections in Eq. (\ref{eq_energy1},\ref{eq_state1},\ref{eq_energy2}).
Because of quantum superposition, the quantum circuit could lead to considerable speedup over classical PT methods.
The framework of our method is presented in Sec.(\ref{Quantum Circuit Implementation}),and in Sec.(\ref{Applications on Extended Hubbard Model}) we will demonstrate the design and optimization of the quantum circuit with the extended Hubbard Model as an example.
In Sec.(\ref{Simulation Results}), we present simulation results conducted in Qiskit. 
The proposed circuit is also implemented on an IBM 27-qubit quantum computer, as presented in Sec.(\ref{Implementation on Quantum Computer}).
Conclusions and discussions are presented in Sec.(\ref{Discussion}).
Additionally, we present analysis on {the} scale and time complexity in Sec.(\ref{Time complexity}) and further applications in Sec.(\ref{Applications}).

\section{Results}
\label{Results}
\subsection{Quantum Circuit Implementation}
\label{Quantum Circuit Implementation}

\begin{figure}[ht]
    \centering
    \includegraphics[width=0.95\textwidth]{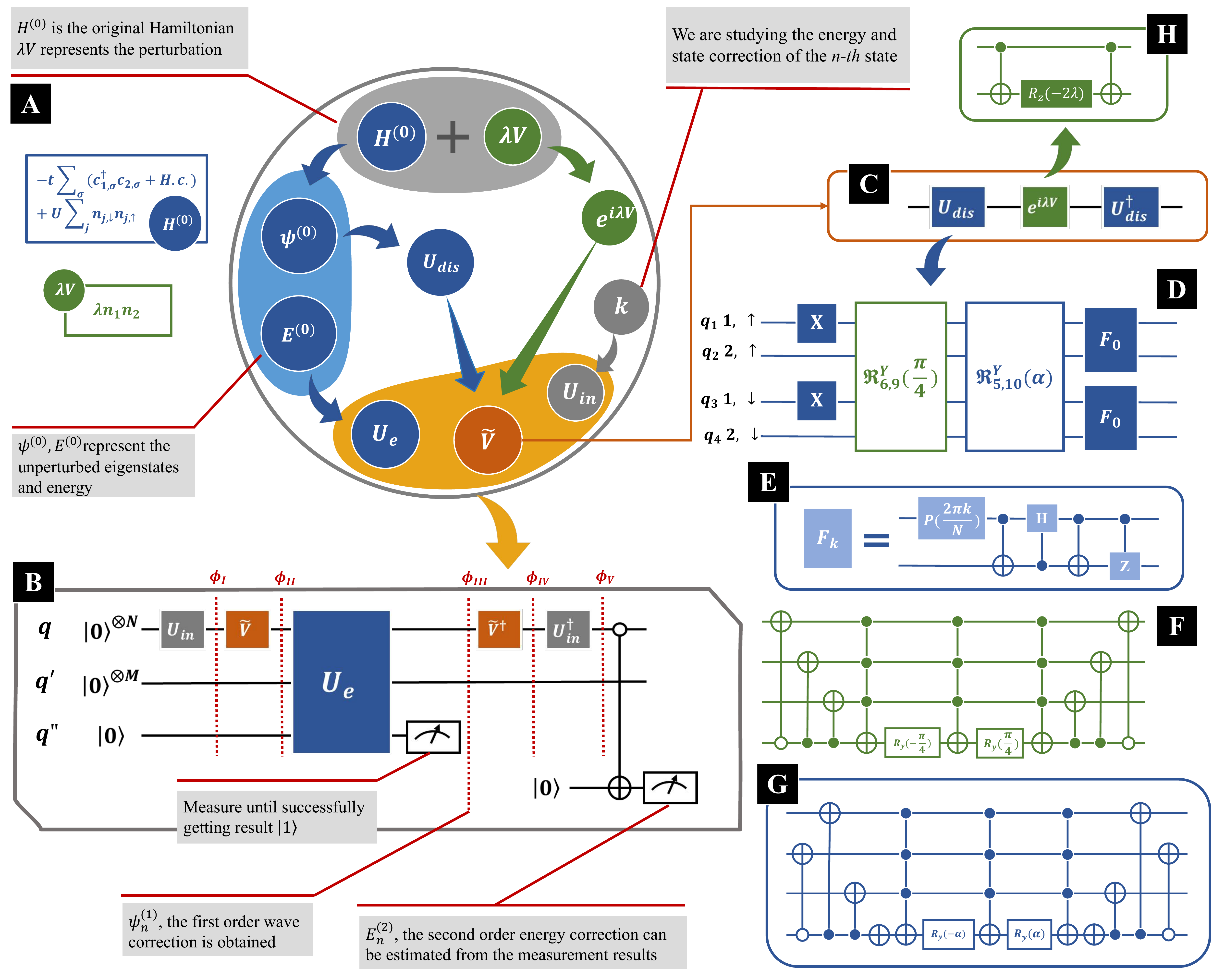}
    \caption{
    {\bf Scheme of the quantum circuit implementation.}
    \\
    (A)Flowchart of the quantum circuit design process.
    %For the given unperturbed Hamiltonian $H^{(0)}$ and perturbation $\lambda V$, we demonstrate how to construct a quantum circuit estimating the first order eigenstate correction $\psi_n^{(1)}$ and the second order energy correction $E_n^{(2)}$, where $n$ indicates that we are studying the energy and state correction of the $n^{th}$ state.
    %Qubits will all be initialized at the ground state $|0\rangle$, and the initializing operator $U_{in}$ will convert them into the $n^{th}$ computational basis, $U_{in}|0\rangle=|n\rangle$.
    %At the beginning, the unperturbed eigen energy $E^{(0)}$ and the corresponding eigenstate $\psi^{(0)}$ can be obtained classically.
    %Based on $E^{(0)}$, $U_e$ is designed estimating $1/(E_n^{(0)}-E_m^{(0)})$ for all possible $m$ values (for $m = n$ the output will be 0).
    %Meanwhile, $U_{dis}$ is designed to convert the computational basis into the unperturbed eigenstates, which guarantees that $U_{dis}|m\rangle = |\psi^{(0)}_m\rangle$.
    %As for the perturbation, since $V$ itself might not correspond to unitary operators, we instead consider $e^{i\lambda V}$ as an approximation.
    %Combining $U_{dis}$ and $e^{i\lambda V}$ together we can get $\Tilde{V}$, which corresponds to the perturbation under computational basis.
    (B)Main structure of the quantum circuit. There are in total $N+M+2$ qubits.
    The first $N$ qubits denoted as $q$ represents the system with $2^N$ basis.
    The next $M$ qubits denoted as $q'$ are included for $U_e$ estimating $1/(E_n^{(0)}-E_m^{(0)})$.
    The others denoted as $q''$ are ancilla qubits for readout.
    $\psi_n^{(1)}$ could be obtained as noted by the dashed line, while $E_n^{(2)}$ would be estimated after measuring the last qubit.
    (C)Structure of $\Tilde{V}$, which contains $U_{dis}$, $U_{dis}^{\dagger}$ and $\exp(i\lambda V)$.
    (D)Structure of $U_{dis}$, where $X$ represents a NOT gate, and $F$ indicates Fourier transformation, as shown in (E).
    There are two special multi-controlled rotation gates $\mathfrak{R}_{6,9}^y(\pi/4)$ and $\mathfrak{R}_{5,10}^y(\alpha)$ in $U_{dis}$, whose structure can be found in  
    (F), (G).
    (H)Quantum circuit simulating the $\exp(i\lambda\sigma^i_z\sigma^j_z)$ term.
    }
    \label{fig_main}
\end{figure}

To estimate the corrections shown in Eq.(\ref{eq_state1},\ref{eq_energy2}) for the $n$-th order terms, there are two important tasks:
{\bf 1.} Estimate the terms of perturbation $\langle \psi_m^{(0)}|V|\psi_n^{(0)}\rangle$;
{\bf 2.} Estimate the inverse of the energy difference term $1/{(E_n^{(0)}-E_m^{(0)})}$ for all $m\neq n$.
Similarly, there are two main modules in our circuit, an operator denoted as $\Tilde{V}$ that simulates the perturbation terms and $U_e$ that estimates the inverse of the energy difference.
A scheme of the quantum circuit estimating the first-order wavefunction correction and second-order energy corrections is presented in Fig.(\ref{fig_main}B).
There are in total $N+M+2$ qubits.
The first $N$ qubits denoted by $q$ represent the system with a basis of size $2^N$.
The next $M$ qubits denoted by $q'$ are ancilla qubits, and the last two are included for readout.
All qubits are initially prepared in the ground state $|0\rangle$.

The first step is to initialize the qubits $q$ into the general state $|k\rangle$, where $|k\rangle$ indicates the corresponding binary form of the state for which we want to calculate corrections.
The grey operator $U_{in}$ in Fig.(\ref{fig_main}B) represents the initializing process, which generally could be fulfilled with simple NOT gates.
For instance, if we would like to study the corrections to the first-excited state,  $U_{in}$ could be a single NOT gate acting on the last qubit of $q$, converting the qubits in $q$ from ground state $|0\rangle$(or $|0\dots 00\rangle$ in binary form) into $|1\rangle$(or $|0\dots 01\rangle$ in binary form).
For simplicity, in the following discussion, we denote the quantum states at certain steps as $|\phi\rangle$, corresponding to the notations in Fig.(\ref{fig_main}B).
After applying $U_{in}$, the qubits are in state $|\phi_I\rangle = |k\rangle_q\otimes|0\rangle_{q'}\otimes|0\rangle_{q''}$, where the subscripts indicate the corresponding qubits, and $k$ indicates that we are studying the corrections for the $k$-th term.

Next, $\Tilde{V}$ is applied on the $q$ qubits.
The perturbation terms $\langle \psi_m^{(0)}|V|\psi_n^{(0)}\rangle$ are approximated with $\langle m|\Tilde{V}|n\rangle$.
The computational basis terms $|n\rangle$  are often different from the original eigenstates $|\psi_n^{(0)}\rangle$ of the unperturbed Hamiltonian $H^{(0)}$.
Consequently, an additional operator $U_{dis}$ is required which converts the computational basis into the original eigenstates, ensuring that $U_{dis}|n\rangle=|\psi_n^{(0)}\rangle$.
The subscript of $U_{dis}$ denotes `disentangle', and $U^{\dagger}_{dis}H^{(0)}U_{dis} = \sum_n{E_n|n\rangle\langle n|}$ is diagonalized under the computational basis.
In Sec.(\ref{Applications on Extended Hubbard Model}) we will present a design of $U_{dis}$, especially for the two-site Hubbard model.
Additionally, a more general quantum circuit implementation of $U_{dis}$ for strongly correlated quantum systems can be found in Ref\cite{verstraete2009quantum}.
If the perturbation $V$ can be exactly decomposed into a sequence of unitary operators, we will have $\Tilde{V} = U^{\dagger}_{dis}V U_{dis}$.
Unfortunately, sometimes $V$ is Hermitian but not unitary.
An alternative is to consider $\exp{(i\lambda V)}$ as an approximation, as $\exp{(i\lambda V)} = I + i\lambda V +\mathcal{O}(\lambda^2)$.
As shown in Fig.(\ref{fig_main}C), the more general design is $\Tilde{V} = U^{\dagger}_{dis}\exp{(i\lambda V)} U_{dis}$, which guarantees that
\begin{equation}
    \langle m|\Tilde{V}|n\rangle
    =
    \delta_{mn} + i\lambda\langle \psi_m^{(0)}|V|\psi_n^{(0)}\rangle
    +\mathcal{O}(\lambda^2)
    \label{eq_v_tilt}
\end{equation}
Here the qubits are converted into the state 
$|\phi_{II}\rangle = \left(\sum_m{\langle m|\Tilde{V}|k\rangle|m\rangle}\right)_q\otimes|0\rangle_{q'}\otimes|0\rangle_{q''}$,
where the state of qubits $q$ is rewritten in the computational basis.

$U_e$ (The blue operator in Fig.(\ref{fig_main}B)) is then applied on $q, q', q''$, generating the inverse of energy difference with
\begin{equation}
    U_e
    \left(
    |n\rangle_q\otimes|0\rangle_{q'}\otimes|0\rangle_{q''}
    \right)
    =
    \left\{
    \begin{aligned}
        &|n\rangle_q\otimes|0\rangle_{q'}\otimes|0\rangle_{q''},
        &n=k
        \\
        &|n\rangle_q\otimes|0\rangle_{q'}\otimes
        \left(
        \sqrt{1-\frac{C^2}{(E_k-E_n)^2}}|0\rangle
        +\frac{C}{E_k-E_n}|1\rangle
        \right)_{q''},
        &n\neq k 
    \end{aligned}
    \right.
    \label{eq_ue}
\end{equation}
where $C$ is a real constant ensuring that $0\leq\left|\frac{C}{E_k-E_n}\right|\leq1$.
$U_e$ contains a few multi controller gates, where the $q$ qubits are control qubits and $q''$ is the target.
$U_e$ is determined by the energy levels, and quantum circuit implementation of $U_e$ is a general method.
More details of $U_e$ can be found in Sec.(\ref{Applications on Extended Hubbard Model}).
Substituting Eq.(\ref{eq_ue}) into $U_e|\phi_{II}\rangle$, the output quantum states become
\begin{equation}
    \begin{split}
        U_e
        |\phi_{II}\rangle
        =&
        \langle k|\Tilde{V}|k\rangle|k\rangle_q
        \otimes|0\rangle_{q'}\otimes|0\rangle_{q''}
        \\
        &+
        \sum_{m\neq k}
        \left\{
        |m\rangle_q
        \otimes|0\rangle_{q'}\otimes
        \left(
        \langle m|\Tilde{V}|k\rangle
        \sqrt{1-\frac{C^2}{(E_k-E_n)^2}}|0\rangle
        +C\frac{\langle m|\Tilde{V}|k\rangle}{E_k-E_n}|1\rangle
        \right)
        \right\}
    \end{split}
\end{equation}
Here the Repeat-Until-Success (RUS)\cite{lim2005repeat} strategy is performed as follows.
Measure the qubit $q''$, and if the readout is $|1\rangle$, then the quantum state will collapse into
$|\phi_{III}\rangle = \sum_{m\neq k}
C'\frac{\langle m|\Tilde{V}|k\rangle}{E_k-E_n}
|m\rangle_q\otimes|0\rangle_{q'}\otimes|1\rangle_{q''}$,
where $C'$ is a normalization constant.
Otherwise, repeat the whole process above until result $|1\rangle$ is obtained when measuring $q''$.
Notice that since $\langle n|\phi_{III}\rangle=C'\langle\psi_n^{(0)}|\psi_k^{(1)}\rangle$,  we now successfully get the first order eigenstate correction.
By measuring $q$ qubits, we can estimate the first order eigenstate $|\psi_k^{(1)}\rangle$.
If we prefer to do further study of $|\psi_k^{(1)}\rangle$ with a quantum circuit, $|\phi_{III}\rangle$ itself is sufficient as an intermediate where the original eigenstates are represented by the corresponding computational basis.
For more demanding requirements, $|\psi_k^{(1)}\rangle$ could be obtained after applying $U_{dis}$ on the qubits $q$, as $U_{dis}|\phi_{III}\rangle=|\psi_k^{(1)}\rangle$.

Since $E_n^{(2)}=\langle\psi_n^{(0)} |V^\dagger|\psi_n^{(1)}\rangle$, we can obtain the second order energy corrections by applying operator $\Tilde{V}^\dagger$ and $U_{in}^\dagger$.
After applying $\Tilde{V}^\dagger$ on qubits $q$ of $|\phi_{III}\rangle$, we have $|\phi_{IV}\rangle=
\sum_{m\neq k}
C'\frac{\langle m|\Tilde{V}|k\rangle}{E_k-E_n}
\Tilde{V}^\dagger|m\rangle_q\otimes|0\rangle_{q'}\otimes|1\rangle_{q''}$.
Then applying $U_{in}^\dagger$, we have $|\phi_{V}\rangle=\sum_{m\neq k}
C'\frac{\langle m|\Tilde{V}|k\rangle}{E_k-E_n}
U_{in}^\dagger\Tilde{V}^\dagger|m\rangle_q\otimes|0\rangle_{q'}\otimes|1\rangle_{q''}$.
Notice that
\begin{equation}
    \begin{split}
        \left(\langle 0|_q\otimes\langle0|_{q'}\otimes\langle1|_{q''}\right)|\phi_V\rangle
        &=
        \sum_{m\neq k}
        C'\frac{\langle m|\Tilde{V}|k\rangle}{E_k-E_n}
        \langle0|U_{in}^\dagger\Tilde{V}^\dagger|m\rangle
        \\
        &=
        \sum_{m\neq k}
        C'\frac{\langle m|\Tilde{V}|k\rangle}{E_k-E_n}
        \langle k|\Tilde{V}^\dagger|m\rangle
        \\
        &=
        \sum_{m\neq k}
        C'\frac{|\langle m|\Tilde{V}|k\rangle|^2}{E_k-E_n}
    \end{split}
\end{equation}
If we measure all $q$ qubits, the probability to get all at state $|0\rangle$ will approximate $E_k^{(2)}$.
Alternatively, a multi-controlled gate could help reduce the measurement times as shown in Fig.(\ref{fig_main}B), where an additional ancilla qubit initialized at $|0\rangle$ is required as the target.

\subsection{Application to the Extended Hubbard Model}
\label{Applications on Extended Hubbard Model}
In this section, we will demonstrate the details of circuit design with the extended Hubbard model.
The Hubbard model is a simple but powerful model of interacting quantum particles in a lattice, which successfully describes the transition between conducting and insulating states\cite{hubbard1963electron}.
The Hamiltonian of the two-site Fermi Hubbard Model is given by
\begin{equation}
    H_{hub} = 
    -t\sum_\sigma{\left(c_{1,\sigma}^\dagger c_{2, \sigma}+c_{2, \sigma}^\dagger c_{1, \sigma}\right)}
    + U\sum_{j =1,2}n_{i,\uparrow}n_{i,\downarrow}
    \label{eq_hubbard}
\end{equation}
where $t$ denotes the transfer integral, $U$ denotes the on-site interaction, and $\sigma =\uparrow,\downarrow$ indicates the spin.
Depending on the atomic species, more general interactions might occur.
A typical example is dipole-dipole interactions induced by polarized dipolar atoms, which is comparatively long-ranged but usually modeled as an interaction between nearest neighbors\cite{micnas1990superconductivity, hofmann2012doublon}.
Adding a dipole-dipole interaction, the Hamiltonian of the extended Hubbard model can be written as\cite{hofmann2012doublon}
\begin{equation}
    H = H_{hub} + W (n_{1,\uparrow}+n_{1,\downarrow})(n_{2,\uparrow}+n_{2,\downarrow})
    \label{eq_pt}
\end{equation}
where $W$ parameterizes the amplitude of dipole-dipole interactions.
When the dipole-dipole interaction is much weaker compared to the hopping term and the on-site interaction, this model becomes a good candidate for PT methods, where $H_{hub}$ is taken as the unperturbed Hamiltonian and the dipole-dipole interaction is regarded as the perturbation. 
Two qubits are required to simulate a single site (spin up and down), so we need in total of four qubits in $q$ to study the two-site extended Hubbard model.
The $q$ qubits are shown in Fig.(\ref{fig_main}D), where $1,2$ indicate the site, $\uparrow,\downarrow$ indicate the spin.
For simplicity, here we study the corrections to the ground state, so that we have $k=0$, and $U_{in} = I$ is the identity operator.

In Fig.(\ref{fig_main}A) we present a flowchart illustrating how to design an appropriate quantum circuit studying the given system with PT methods.
We start with the unperturbed Hamiltonian $H^{(0)}$ and the perturbation term $\lambda V$, where $\lambda\ll 1$.
The first step is to derive the eigenenergy $E_n^{(0)}$ and corresponding eigenstates $\psi_n^{(0)}$ of $H^{(0)}$.
$H_{hub}$ being a typical model well-developed in the past 50 years, the eigenenergies and eigenstates can be regarded as known terms.
%As for other models, one could calculate the eigen energy and eigenstates with the classical method, while the Quantum Phase Estimation Algorithm would also be an option if the quantum method is preferred.
With $\psi_n^{(n)}$, we can design $U_{dis}$ that converts the computational basis $|n\rangle$ into the corresponding eigenstate, as $U_{dis}|n\rangle=|\psi_n^{(0)}\rangle$.
In addition, we can design $U_{e}$ generating the inverse of the energy difference.
In Fig.(\ref{fig_main}A), these terms are all colored blue, as the operators $U_{e}$ and $U_{dis}$ are only determined by the unperturbed Hamiltonian $H^{(0)}$.
In other words, provided that a new perturbation is applied on the same $H^{(0)}$, these operators could be kept without any changes.
Regarding the perturbation term, we consider $\exp(i\lambda V)$ since the dipole-dipole interaction cannot be decomposed exactly into a sequence of unitary operators.
% As for the perturbation term, since the dipole-dipole interaction cannot be exactly decomposed into a sequence of unitary operators, we here consider $\exp(i\lambda V)$ instead.
Implementation of these key operators is as follows.
\newline

{\noindent\bf Implementation of $U_{dis}$}

Fig.(\ref{fig_main}D) is a schematic of the operator $U_{dis}$.
%To diagonalize the hopping term $c_{1,\sigma}^\dagger c_{2, \sigma}+H.c$, we apply a quantum Fourier transform (QFT) on $q_1$ and $q_2$ for spin up and $q_3$ and $q_4$ for spin down. The QFT is constructed as shown in Figure (\ref{fig_main}E), where $P$ denotes the phase gate.
Noticing that a Fourier Transform can diagonalize the hopping term $c_{1,\sigma}^\dagger c_{2, \sigma}+H.c$, we apply a quantum Fourier transform (QFT) on $q_1, q_2$ (spin up) and $q_3, q_4$ (spin down); the construction of the QFT can be found in Fig.(\ref{fig_main}E), where $P$ represents the phase gate.
Due to the existence of on-site interactions, QFT itself is not yet sufficient.
Two additional operators denoted as $\mathfrak{R}_{6,9}^y(\pi/4)$ and $\mathfrak{R}_{5,10}^y(\alpha)$ are required, which act as special multi-controlled rotation gates.
The matrix form of $\mathfrak{R}_{5,10}^y(\alpha)$ is
\begin{equation}
    \mathfrak{R}_{5,10}^y(\alpha)
    =
    \begin{array}
    {@{}r@{}c@{}c@{}c@{}c@{}c@{}c@{}c@{}c@{}l@{}}
    &1 &2  &\cdots &5 &\cdots &10 &\cdots &16\\
    \left.\begin{array}
    {c} 1 \\ 2  \\ \vdots \\ 5 \\ \vdots \\ 10 \\ \vdots \\16 \end{array}
    \right(
        & \begin{array}{c} 1 \\ 0 \\ \vdots \\ 0 \\ \vdots \\ 0 \\ \vdots \\0\end{array}
        & \begin{array}{c} 0 \\ 1 \\ \vdots \\ 0 \\ \vdots \\ 0 \\ \vdots \\0\end{array}
        & \begin{array}{c} \cdots \\ \cdots \\ \ddots \\ \quad \\ \quad \\ \quad \\ \quad \\ \cdots\end{array}
        & \begin{array}{c} 0 \\ 0 \\ \vdots \\ \cos\alpha \\ \vdots \\ \sin\alpha \\ \vdots \\0\end{array}
        & \begin{array}{c} \cdots \\ \cdots \\ \qquad \\ \qquad \\ \ddots \\ \qquad \\ \qquad \\ \cdots\end{array}
        & \begin{array}{c} 0 \\ 0 \\ \vdots \\ -\sin\alpha \\ \vdots \\ \cos\alpha \\ \vdots \\0\end{array}
        & \begin{array}{c} \cdots \\ \cdots \\ \qquad \\ \qquad \\ \qquad \\ \qquad \\ \ddots \\ \cdots\end{array}
        & \begin{array}{c} 0 \\ 0 \\ \vdots \\ 0 \\ \vdots \\ 0 \\ \vdots \\1\end{array}
        & \left)\begin{array}{c} \\ \\ \\ \\ \\ \\ \quad\\ \quad\\  \quad\\  \end{array}\right.
  \end{array}
  \label{V_so}
\end{equation}
where the numbering of the columns and rows indicates the corresponding eigenstates, and
\begin{equation}
    \alpha = -2\arccos\left(\frac{2t +\sqrt{ U^2 / 4 + 4t^2}}{\sqrt{U^2/4 +(2t + \sqrt{ U^2 / 4 + 4t^2})^2}}\right)
\end{equation}
Implementation of these two special operations can be found in Fig.(\ref{fig_main}F,\ref{fig_main}H).
%{\color{red} Most likely you mean Fig.1F, 1G as 1H seems to be implementation of $e^{iV}$ which is not these rotation operators.}
%$\mathfrak{R}_{6,9}^y(\pi/4)$ has the similar matrix form. 
Additionally, there are two NOT gates applied on $q_1, q_3$, which are included to ensure that the state $|0\rangle$ (or $|0000\rangle$ in binary form) corresponds to the ground state $|\psi_0^{(0)}\rangle$.
\newline

{\noindent\bf Implementation of $\exp(i\lambda V)$}

Using Jordan-Wigner transformation\cite{batista2001generalized}, $\sigma_z = 1-2n$, the perturbation term in Eq.(\ref{eq_pt}) can be written as
\begin{equation}
    \lambda V = \frac{W}{4}
    (\sigma^z_{1,\uparrow}\sigma^z_{2,\uparrow}
    +\sigma^z_{1,\uparrow}\sigma^z_{2,\downarrow}
    +\sigma^z_{1,\downarrow}\sigma^z_{2,\uparrow}
    +\sigma^z_{1,\downarrow}\sigma^z_{2,\downarrow})
    +\frac{W}{4}
    -\frac{W}{2}
    (n_{1,\uparrow}+n_{2,\uparrow}+n_{1,\downarrow}+n_{2,\downarrow})
\end{equation}
In the Hubbard model, the conservation of the total number of particles implies that the last term is a constant, leaving only the first term as the non-trivial component.
% The total number of particles is conserved in the Hubbard model, so that the last term is a constant, and only the first term is non-trivial.
For simplicity, we denote $\lambda = W/4\ll1$.
With first-order Trotter decomposition\cite{lloyd1996universal}, we have
\begin{equation}
    \exp(i\lambda V) = 
    \exp\left(i\lambda\sigma^z_{1,\uparrow}\otimes\sigma^z_{2,\uparrow}\right)
    \dot\exp\left(i\lambda\sigma^z_{1,\uparrow}\otimes\sigma^z_{2,\downarrow}\right)
    \dot\exp\left(i\lambda\sigma^z_{1,\downarrow}\otimes\sigma^z_{2,\uparrow}\right)
    \dot\exp\left(i\lambda\sigma^z_{1,\downarrow}\otimes\sigma^z_{2,\downarrow}\right)
\end{equation}
The quantum circuit simulating $\exp(i\lambda \sigma^z_j\otimes\sigma^z_k), (j\neq k)$ is presented in Fig.(\ref{fig_main}H), and more details can be found in ref\cite{whitfield2011simulation}.
\newline

{\noindent\bf Implementation of $U_e$}

\begin{figure}[ht]
    \centering
    \includegraphics[width=0.95\textwidth]{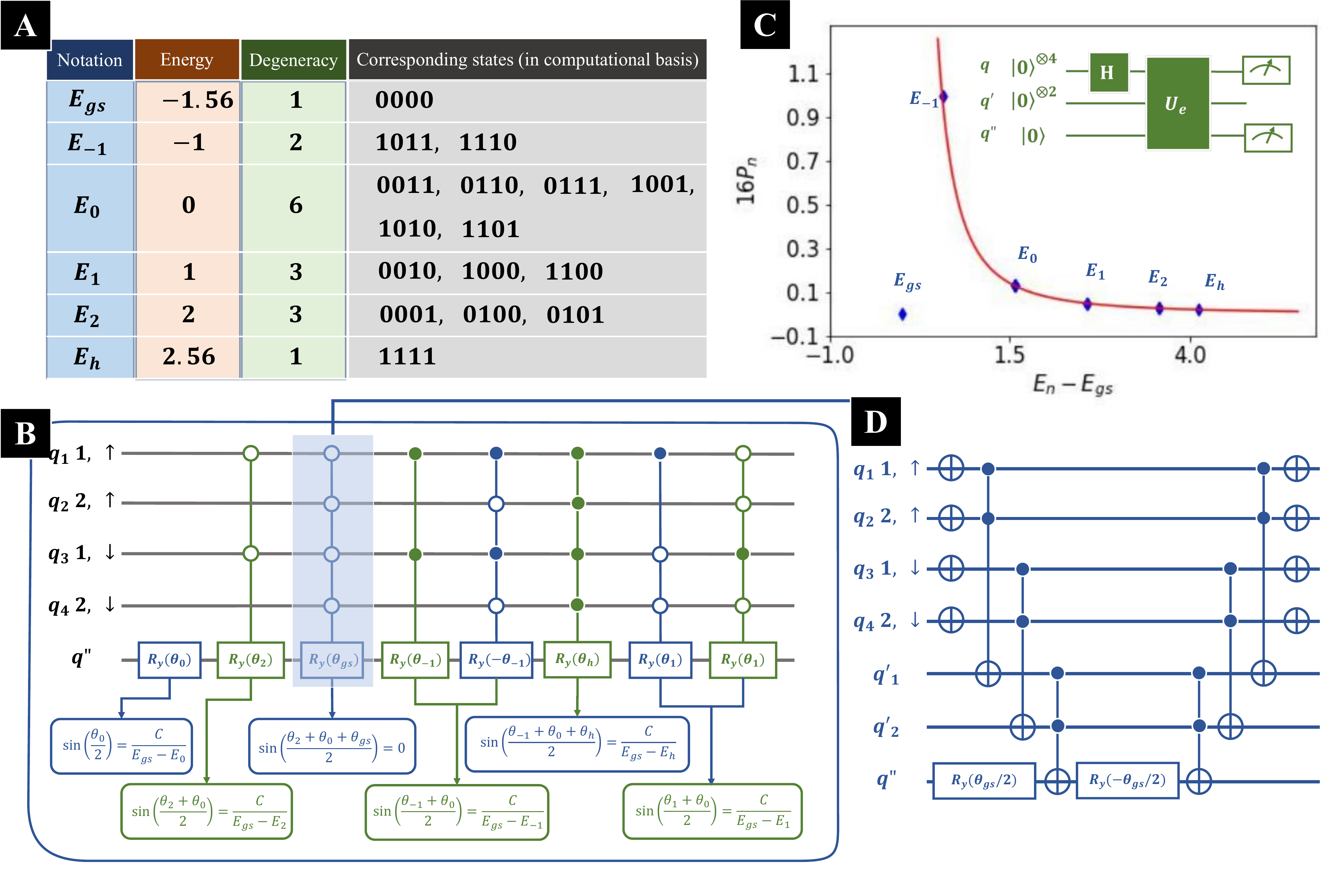}
    \caption{
    {\bf Unperturbed energy levels and implementation of $U_e$.}
    (A)Table of the energy levels, degeneracy, and corresponding states under the computational basis for the unperturbed Hamiltonian $H^{(0)}$, where for simplicity, we set $t=1$, $U=1$.
    (B)Quantum circuit implementation of $U_e$ (The operator $U_e$ in Fig.(\ref{fig_main}B)), where the ancilla qubits $q'$ are not plotted.
    $U_e$ is constructed with mainly multi-controlled rotation gates.
    All the parameters $\theta$ are determined by the energy levels of the unperturbed Hamiltonian.
    (C)Calibration of $U_e$.
    The calibration circuit is plotted in green, shown in the right-upper corner.
    In the calibration circuit, all qubits are initialized as $|0\rangle$.
    Hadamard gates are then applied on each $q$ qubit, preparing the $q$ qubits in a uniform superposition.
    Then $U_e$ is applied and qubits $q$ and $q'$ are measured.
    We denote the probability to find the qubits $q$ at state $|n\rangle$ and $q''$ at state $|1\rangle$ as $P_n$.
    The X-axis denotes the energy difference $E_n-E_{gs}$, while the Y-axis denotes $16P_n$.
    (D)Decomposition of the multi-controlled rotation gate with light blue background shown in (B) (The second multi-controlled gate from left, colored in blue).
    }
    \label{fig_ue}
\end{figure}

Before discussing the construction of $U_e$, we need to first calculate the unperturbed energy levels.
For simplicity, here we set $t=1$, $U=1$.
Energy levels, degeneracy, and corresponding states under the computational basis for the two-site Hubbard model $H^{(0)}$ are presented in Fig.(\ref{fig_ue}A).
Here the eigenstates and eigenenergies are all included (4 for the one-electron sector, 6 for the two-electron sector, 4 for the three-electron sector, and 2 trivial terms: 4-electron and 0-electron).
The ground state energy is denoted as $E_{gs}$, while $E_h$ represents the energy of the highest excited state.
These two states correspond to the ground state and highest excited state of the 2-electron Hubbard model (the half-filled case of strong correlations).
$E_{0, \pm1, 2}$ denote the other excited state energies, where the subscripts denote the corresponding energy.
Fig.(\ref{fig_ue}B) is the quantum circuit implementation of $U_e$, where the ancilla qubits $q'$ are not plotted.
$U_e$ is constructed with mainly multi-controlled rotation gates, where the dot on the control qubit indicates that the rotation gate works when this control qubit is $|1\rangle$, and the circle on the control qubit indicates that the rotation gate works when this control qubit is $|0\rangle$.
First the energy level $E_0$ is considered as the `default value', as it has the most degeneracy.
Hence a simple Ry gate is applied directly on $q''$, leading to $\sin(\theta_0/2)=C/(E_{gs}-E_0)$, where $C$ is the constant in Eq.(\ref{eq_ue}).
Then we study $E_2$, which contains three degenerate states corresponding to $|0001\rangle$, $|0100\rangle$, and $|0101\rangle$ in the computational basis.
Notice that all three states share the same first and third digit as 0, so that a multi-controlled gate with $q_1$ and $q_3$ as control qubit is applied, leading to $\sin((\theta_0+\theta_2)/2)=C/(E_{gs}-E_2)$.
Now the first multi-controlled rotation gate from left (colored in green) in Fig.(\ref{fig_ue}B) is constructed.
In fact, there is an additional state $|0000\rangle$ sharing the first and third digits as 0, which corresponds to the ground state.
We are now studying the corrections to the ground state, and we need to insure $q''$ is always at state $|0\rangle$ when the control qubits are at state $|0000\rangle$.
Therefore, the second multi-controlled rotation gate from left (colored in blue) in Fig.(\ref{fig_ue}B) is constructed, ensuring that $\sin((\theta_0+\theta_2+\theta_{gs})/2)=0$.
The decomposition of this multi-controlled gate is presented in Fig.(\ref{fig_ue}D), which contains only NOT gates, single qubit Ry gates, and Toffoli gates.
Similarly, the other multi-controlled gates can be constructed, with the corresponding parameters $\theta$ determined by the energy levels.
In Fig.(\ref{fig_ue}C) we present the calibration of $U_e$.
%\textcolor{red}{The calibration circuit, located in the upper-right corner and plotted in green, initializes all qubits to the $|0\rangle$ state.}
The calibration circuit is plotted in green, shown in the right-upper corner.
In the calibration circuit, all qubits are initialized as $|0\rangle$.
Hadamard gates are then applied on each $q$ qubit, preparing the $q$ qubits in a uniform superposition.
Then $U_e$ is applied and qubits $q$ and $q'$ are measured.
{Let $P_n$ denote the probability of finding the qubits $q$ in state $|n\rangle$ and $q''$ in state $|1\rangle$.}
% Denote the probability to find the qubits $q$ at state $|n\rangle$ and $q''$ at state $|1\rangle$ as $P_n$.
Theoretically, $P_n=\frac{1}{16} \cdot \frac{C^2}{(E_{gs}-E_n)^2}$. The
X-axis represents the energy difference $E_n-E_{gs}$, while the Y-axis denotes $16P_n$.
The red curve represents the ideal result, while the blue dots are simulation results for each energy level.
%{To maximize $P_{-1}$, we set $C$ equal to $1/(E_{gs}-E_{-1})$.}
Here we set $C = 1/(E_{gs}-E_{-1})$, so that $P_{-1}=1$ reaches the maximum.
Calibration results in Fig.(\ref{fig_ue}C) prove the ability of $U_e$ shown in Fig.(\ref{fig_ue}B) to generate the inverse of energy differences.

\subsection{Simulation Results}
\label{Simulation Results}

\begin{figure}[ht]
    \centering
    \includegraphics[width = 0.95\linewidth]{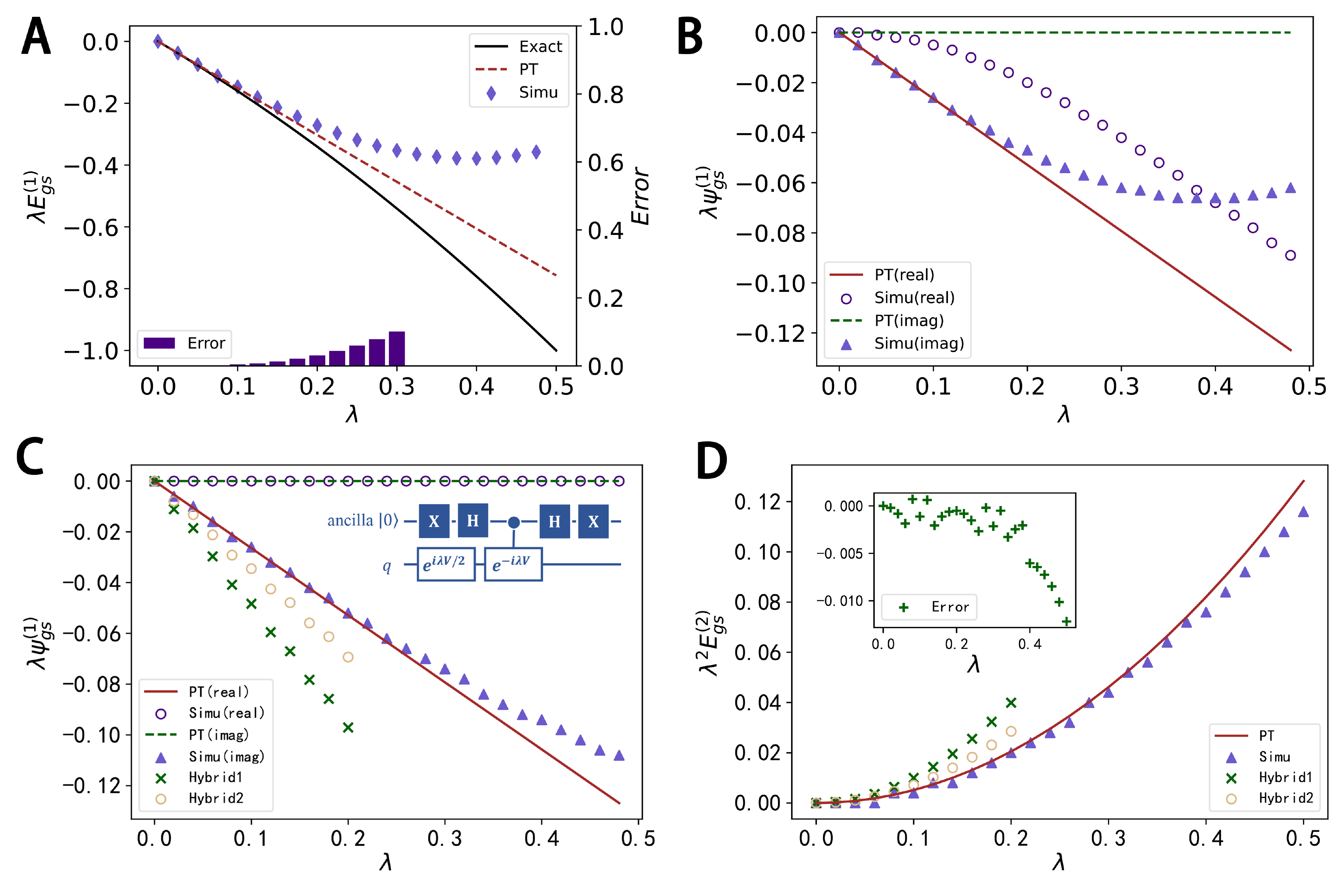}
    \caption{
    {\bf Simulation results, experiment results, and the corresponding prediction by PT methods.}
    All the simulations are performed with Qiskit, while the experiments are implemented on `ibmq\_montreal'.
    %We studied the first and second order energy correction, and first order eigenstate correction for the ground state of the extended Hubbard model as shown in Eq.(\ref{eq_hubbard}) and Eq.(\ref{eq_pt}), where we set $t=1$, $U=1$ for simplicity. 
    (A) First order energy correction $\lambda E_{gs}^{(1)}$.
    The black curve represents the exact energy change $E_{gs}-E_{gs}^{(0)}$, while the brown dashed line is the first order energy correction $\lambda E_{gs}^{(1)}$.
    The purple markers denote the estimation of $\lambda E_{gs}^{(1)}$ with the quantum circuit.
    Purple bars at the bottom denote the error between the first-order PT energy correction $\lambda E_{gs}^{(1)}$ and the estimation with the quantum circuit.
    %For $\lambda<0.1$, the simulation results fit well with the PT prediction, while for greater $\lambda$, both the PT prediction and the estimation on quantum circuit are not capable to approximate the exact energy change.
    (B,C) First order eigenstate correction $\lambda|\psi_{gs}^{(1)}\rangle$.
    The brown lines denote the prediction based on PT methods (solid line for the real part and dashed line for the imaginary part).
    Purple markers denote the estimation with the quantum circuit (triangles for the imaginary part and circles for the real part).
    %In (B) we apply $exp(i\lambda V)$ to approximate the perturbation, while in (C) $exp(i\lambda V/2)-exp(-i\lambda V/2)$ is applied to approximate the perturbation.
    The improved circuit is shown in the upper right of (C).
    (D) Second order energy correction $\lambda^2 E_{gs}^{(1)}$.
    The brown curve represents the prediction of PT, while the purple triangles denote the estimation with the quantum circuit.
    The error between the quantum estimation and the PT prediction is presented in the upper left of (D).
    In (C) and (D), the PT corrections estimated from the hybrid calculations are also presented, as marked with green cross symbols (Hybrid 1) and orange circles (Hybrid 2).
    In Hybrid 1 we applied the full steps of $U_e$, but in Hybrid 2 we only kept the main terms with minimum multi-controller gates.
    }
    \label{fig_simu}
\end{figure}

We studied the first and second-order energy corrections, and first-order eigenstate correction for the ground state of the extended Hubbard model as shown in Eq.(\ref{eq_hubbard}) and Eq.(\ref{eq_pt}), where we set $t=1$, $U=1$ for simplicity and the simulation results performed on Qiskit are presented in Fig.(\ref{fig_simu}). 

Fig.(\ref{fig_simu}A) shows the approximation of first order energy correction $\lambda E_{gs}^{(1)}$.
According to Eq.(\ref{eq_v_tilt}), we have ${\bf Im}(\langle n|\Tilde{V}|n\rangle)=\lambda E^{(1)}_n +\mathcal{O}(\lambda^2)$.
Thus the first order energy correction can be approximated by the estimation of ${\bf Im}(\langle m|\Tilde{V}|n\rangle)$, requiring only the $q$ qubits implementing the operator $\Tilde{V}$.
In Fig.(\ref{fig_simu}A), the black curve represents the exact energy change $E_{gs}-E_{gs}^{(0)}$, while the brown dashed line is the first order energy correction $\lambda E_{gs}^{(1)}$.
The purple markers denote the estimation of $\lambda E_{gs}^{(1)}$ with the quantum circuit.
The purple bars at the bottom denote the error between the first-order PT energy correction $\lambda E_{gs}^{(1)}$ and the estimation with the quantum circuit.
For $\lambda<0.1$, the simulation results fit well with the PT prediction, while for greater $\lambda$, both the PT prediction and the estimation on the quantum circuit do not do well in approximating the exact energy change.

{We show the study of the first-order eigenstate correction $\lambda|\psi_{gs}^{(1)}\rangle$ in Fig.(\ref{fig_simu}B,C).}
The brown lines denote the prediction based on PT methods (solid line for the real part and dashed line for the imaginary part).
Purple markers denote the estimation with the quantum circuit (triangles for the imaginary part and circles for the real part).
In Fig.(\ref{fig_simu}B) we apply $exp(i\lambda V)$ to approximate the perturbation.
According to Eq.(\ref{eq_v_tilt}), a global phase $-i$ is included in the first order term; thus the imaginary part of the output will approximate $|\psi_{gs}^{(1)}\rangle$.
For $\lambda<0.1$, the simulation results fit well with the PT prediction, while for greater $\lambda$, the real part of the simulation result increases rapidly, which corresponds to the $\mathfrak{\lambda^2}$ term in Eq.(\ref{eq_v_tilt}).
{To approximate the perturbation and eliminate the $\mathfrak{\lambda^2}$ term, we apply $exp(i\lambda V/2)-exp(-i\lambda V/2)$ as shown in Figure (\ref{fig_simu}C). The improved circuit can be found in the upper-right corner of the same figure.}
% To eliminate the $\mathfrak{\lambda^2}$ term, in Fig.(\ref{fig_simu}C) $exp(i\lambda V/2)-exp(-i\lambda V/2)$ is applied to approximate the perturbation.
% The improved circuit is shown in the upper right of Fig.(\ref{fig_simu}C).
%\textcolor{red}{Let us assume that the $q$ qubits are initially prepared at $|\psi_{input}\rangle$. If the ancilla qubit is measured and yields the result $|1\rangle$, the $q$ qubits will be in the state $(exp(i\lambda V/2)-exp(-i\lambda V/2))|\psi_{input}\rangle$.}
Assume that the $q$ qubits are initially prepared at $|\psi_{input}\rangle$.
If the ancilla qubit is measured and the result is $|1\rangle$, then we have the $q$ qubits at state $(exp(i\lambda V/2)-exp(-i\lambda V/2))|\psi_{input}\rangle$.

In Fig.(\ref{fig_simu}D) we study the second order energy correction $\lambda^2 E_{gs}^{(2)}$.
Similarly, $exp(i\lambda V/2)-exp(-i\lambda V/2)$ is applied to approximate the perturbation.
The brown curve represents the prediction of PT, while the purple triangles denote the estimation with the quantum circuit.
The error between the quantum estimation and the PT prediction is presented in the upper left of Fig.(\ref{fig_simu}D).
As $exp(i\lambda V/2)-exp(-i\lambda V/2)$ is applied to approximate the perturbation, the $\lambda/2$ terms instead of $\lambda$ itself dominate the convergence, so that in Fig.(\ref{fig_simu}C, D) the simulation results fit well with the PT prediction for $\lambda<0.2$.
Therefore, we can collect the results for a range of $\lambda$ values, and then derive $|\psi_{gs}^{(1)}$, $\rangle$ $E_{gs}^{(2)}$ with linear regression methods.

\subsection{Implementation on a Quantum Computer}
\label{Implementation on Quantum Computer}

In addition to the simulation performed in Qiskit,  we also implement the proposed circuit on IBM's quantum hardware.
As discussed in Sec.(\ref{Applications on Extended Hubbard Model}),
there are three key operations in our proposed circuit, $U_{dis}$, $exp(i\lambda V)$, and $U_e$.
Both $U_{dis}$ and $exp(i\lambda V)$ only act on the first four qubits $q_{1,2,3,4}$.
With the typical Trotter decomposition\cite{lloyd1996universal}, $exp(i\lambda V)$ could be implemented with a few simple CNOT gates and Rz gates.
In addition, although there are two complicated gates $\mathfrak{R}_{6,9}^y(\pi/4)$ and $\mathfrak{R}_{5,10}^y(\alpha)$ in $U_{dis}$, they could be replaced by a two-qubit Bogoliubov transformation along with quantum Fourier transformation, as discussed in Ref\cite{verstraete2009quantum}.
{Implementing $U_e$ on quantum devices during the NISQ era is particularly challenging since it involves multiple multi-controller gates and acts on all of the qubits $q, q', q''$. In comparison to the other two operations, $U_e$ is considerably more complex to implement.}
% However, $U_e$ acts on all of the qubits $q, q', q''$, and contains several multi-controller gates.
% Compared with the other two operations, it is much more challenging to implement $U_e$ on quantum devices in the NISQ era.
In this section, we will concentrate on the implementation of $U_e$ on a quantum computer.

\begin{figure}[ht]
    \centering
    \includegraphics[width=0.95\textwidth]{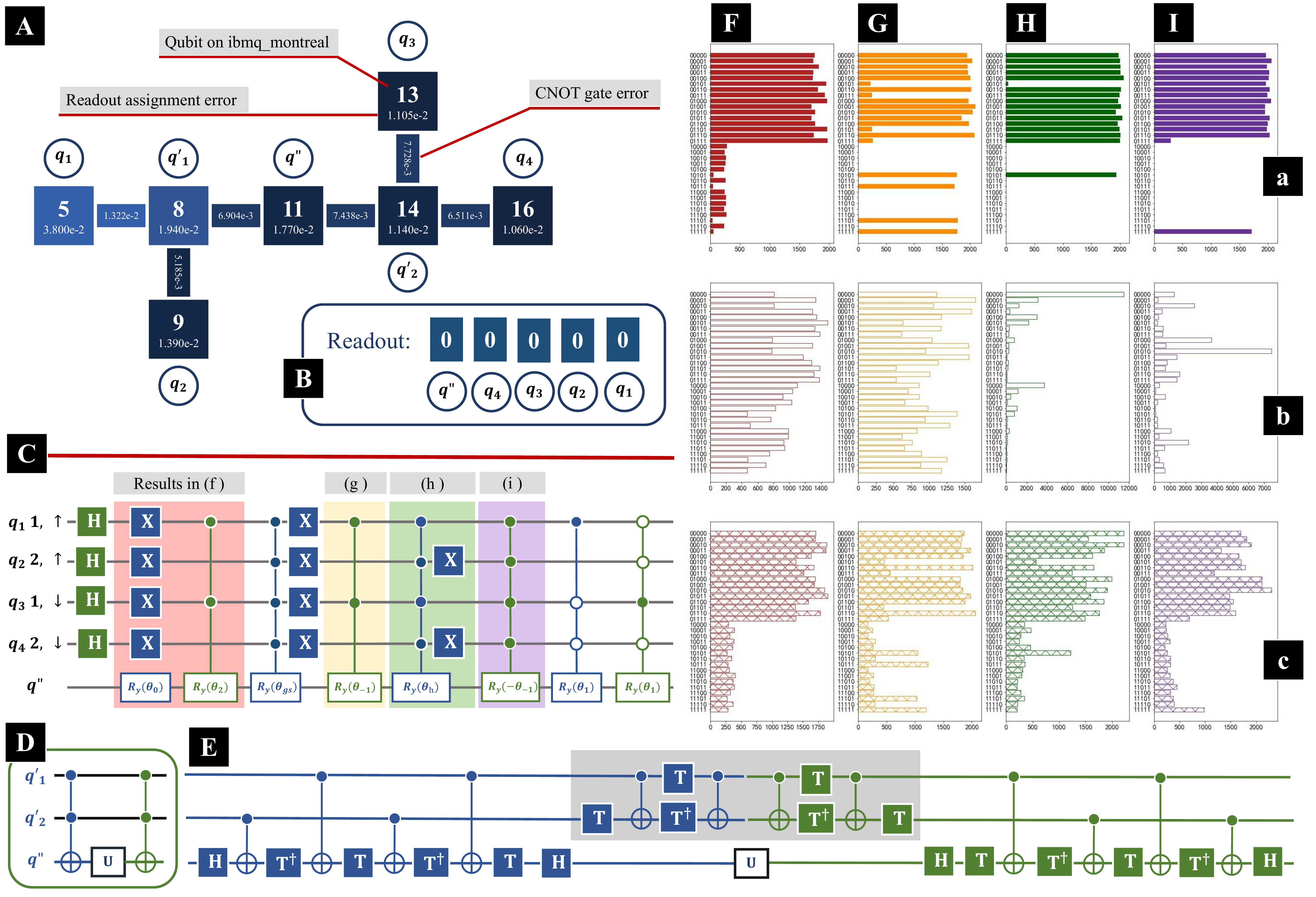}
    \caption{
    {\bf Implementation of $U_e$ on quantum computer `ibmq\_montreal'.}
    (A) Qubit mapping on `ibmq\_montreal', each square indicating a qubit on the quantum computer.
    (B) Relationship between the digits in readout and original qubits.
    (C) Structure of $U_e$. Here we study parts of $U_e$ separately, with results of each part presented in (F, G, H, I), wherein the first row we present the simulation results, in the second row we present the original result, and in the last row we present the output with improved CCNOT gates pairs, as shown in (D, E).
    (D) A pair of CCNOT gates between $q'_{1,2}, q''$ on either side of an operator denoted as $U$ acting on the target qubit $q''$.
    (E) The decomposition of operations is shown in (D). As the operations with grey background cancel out and can be excluded, there is no operation between $q'_{1,2}$, which could avoid several SWAP gates in the multi-controller gates.
    }
    \label{fig_ibmq}
\end{figure}

There are in total 7 qubits involved in $U_e$: $q_{1,2,3,4}$ representing the physical system, $q'_{1,2}$ included to construct the multi-controller gates, and $q''$ for readout.
In Fig.(\ref{fig_ibmq}C), we present the structure of $U_e$.
Due to $U_e$ being a complicated operation, we study the contribution of the multi controller gates separately, and the parts of $U_e$ are applied individually.
Initially, all qubits are initialized to the ground state $|0\rangle$.
Then Hadamard gates convert $q_{1,2,3,4}$ into a uniform superposition.
Next, part of $U_e$ is applied, and $q_{1,2,3,4}$ along with $q''$ are measured at the end, resulting in a binary number.
The relationship between the digits in readout and original qubits is presented in Fig.(\ref{fig_ibmq}B). 
In particular, here we present the results of four typical parts in $U_e$.
Two of them mainly contain 2-controller-rotation gates, which are the operations with background colored in light red and light yellow in Fig.(\ref{fig_ibmq}C), and the corresponding results are presented in Fig.(\ref{fig_ibmq}F,G).
The other two parts mainly contain 4-controller-rotation gates, as the operations with background colored in light green and light purple in Fig.(\ref{fig_ibmq}C), whose contribution can be found in Fig.(\ref{fig_ibmq}H,I).
In the first row of Fig.(\ref{fig_ibmq}F, G, H, I), we present the ideal result without errors, which is obtained from IBM's simulator, named `simulator\_statevector'.
There are 32000 shots in each job (the same as in the following jobs on the real quantum computer).
We then ran the parts of $U_e$ on IBM's 27-qubit quantum computer `ibmq\_montreal', and the results can be found in the second row of Fig.(\ref{fig_ibmq}F, G, H, I).
At this stage, the bare uncorrected results shown in Fig.(\ref{fig_ibmq}F, G, H, I), first row are far different from expectations, with not even the bare shape recognizable.

% When a two-qubit operation is performed between the qubits without a physical connection, auxiliary operations such as SWAP gates are required to form connections between these qubits, which unavoidably contribute to additional errors.
% It is necessary to optimize the qubit mapping by minimizing the required amount of auxiliary operations.
{When a two-qubit operation is performed between the qubits without a physical connection, auxiliary operations such as SWAP gates are required, which unavoidably contribute to additional errors. It is necessary to optimize the qubit mapping by minimizing the required amount of auxiliary operations.}
Here we pick seven qubits on `ibmq\_montreal' to implement $U_e$, as shown in Fig.(\ref{fig_ibmq}A), where each square indicates a qubit on the quantum computer, the nearby circle infers the qubit mapping and neighbor qubits are connected.
The readout assignment error of each qubit is presented at the bottom of each square, while the CNOT gate error is presented on the bar connecting the squares.
The errors usually change after calibration.
When we ran the jobs, the median CNOT gate error of `ibmq\_montreal' was $8.636e-3$, and the median readout error is $1.410e-2$.

As presented in Fig.(\ref{fig_ue}D), the 4-controller-rotation gate can be decomposed into CCNOT gates (Toffoli gates) and single qubit rotation gates.
In Fig.(\ref{fig_ibmq}D), we plot a pair of CCNOT gates between $q'_{1,2}, q''$, bracketing the operator denoted as $U$ acting on the target qubit $q''$.
The left CCNOT gate (colored in blue) can be decomposed into several single qubit gates and CNOT gates, as shown in the left part (colored in blue) of Fig.(\ref{fig_ibmq}E).
In the original decomposition, there are not only CNOT gates between $q'_1$ and $q''$, $q'_2$ and $q''$, but also CNOT gates between $q'_1$ and $q'_2$.
However, on the quantum computer there is no direct connection between $q'_{1,2}$, as shown in Fig.(\ref{fig_ibmq}A), and several quantum SWAP gates are required, which can lead to considerable error as shown in the second row of Fig.(\ref{fig_ibmq}F, G, H, I).
Luckily, the inverse of a CCNOT gate is itself, so we can decompose the other CCNOT gate as the inverse, as shown in the right part (colored in green) of Fig.(\ref{fig_ibmq}E).
Notice that the operations with light grey background cancel out and can be excluded, and there are no more CNOT gates between $q'_{1,2}$.
Similarly, we can decompose the other CCNOT gate pairs in Fig.(\ref{fig_ue}D).

In the final decomposition of the 4-controller-rotation gate, there are only CNOT gates between the neighbors, and no auxiliary SWAP gate is required.
On the other hand, the two-controller-rotation gate can be decomposed into four CNOT gates, two CCNOT gates, and two single qubit rotation gates (similar to the decomposition in Fig.(\ref{fig_ue}D), but replace the first two and last two CCNOT gates with CNOT gates), where no auxiliary SWAP gate is required either.
%In addition of the qubit mapping techniques, variational rotation gates could be introduced on $q,q'$ between the multi-controller gates, ensuring that the control qubits $q$ are still at the uniform superposition state, while $q'$ are still at state $|0\rangle$.
In the third row of Fig.(\ref{fig_ibmq}F, G, H, I), we present the results on `ibmq\_montreal' with the improved qubit mapping techniques.
%Though still imperfect, the final output is much more accurate compared with the original.
Additionally, due to the degeneracy in unperturbed energy levels, it is possible to reduce the number of multi-controller gates in $U_e$.
Generally, $U_e$ as shown in Fig.(\ref{fig_ibmq}C) is equivalent to decomposition with 1 Ry gate, 4 CRy gates, 6 CCRy gates, 4 CCCRy gates, and 1 CCCCRy gate (a CRy gate contains one controller qubit, a CCRy gate contains two and so on), some of which could be excluded due to the degeneracy in unperturbed energy levels.
Moreover, when studying $U_e$ on the quantum computer, we notice that some multi-controller Ry gates with a small parameter are extremely sensitive, and the magnitude of their contribution is less than their average error.
In our experiment, multi-controller gates with $\theta_{gs}$ and ${\theta_1}$ lead to more error than contribute to the overall result.
In the supplementary materials, a detailed discussion of $U_e$ can be found.

We estimate the PT corrections with a hybrid calculation.
The contributions of $U_e$ are estimated on the quantum computer separately, while the operations $U_{dis}$ and $exp(i\lambda V)$ are estimated via the simulator classically.
In Figure (\ref{fig_simu}C, D) we present the PT corrections estimated from the hybrid calculations, as marked with green cross symbols (Hybrid 1) and the orange circles (Hybrid 2).
In Hybrid 1 we applied the full steps of $U_e$, but in Hybrid 2 we only kept the main terms with minimum multi-controller gates (terms with small magnitude/large error $\theta_{gs}$ and ${\theta_1}$ are excluded).
Compared with Hybrid 1, Hybrid 2 results are much closer to the simulation results.
The first-order eigenstate correction is estimated from the measurement results of $q$ and $q''$, containing the contributions of both the imaginary part and the real part.
Here we concentrate on the PT eigenstate and eigenenergy corrections for the ground state, where the main contribution is from the term proportional to $\frac{1}{E_h-E_{gs}}$, where $E_h$ is the state with the highest energy.
As shown in Fig.(\ref{fig_ue}C), $\frac{1}{E_h-E_{gs}}$ is the minimum among all such energy difference-dependent terms. 
%\textcolor{red}{Due to its sensitivity to errors, the corresponding output in $U_e$ is relatively small, leading to PT corrections from both hybrid calculations that exceed the simulation results (indicated by purple triangles in Fig.(\ref{fig_simu}C, D)). The use of state-of-the-art quantum error mitigation techniques\cite{temme2017error, li2017efficient} could further enhance the accuracy of the results.}
Consequently, the corresponding output in $U_e$ is quite small and sensitive to the existence of errors.
Thus, the PT corrections from both hybrid calculations are much greater than the simulation results, which are marked with purple triangles in Fig.(\ref{fig_simu}C,D).
Even more accurate results could be obtained with state-of-the-art quantum error mitigation techniques\cite{temme2017error, li2017efficient} .

%It is still extremely challenging to implement the whole circuit on a quantum computer with current CNOT gate error rates.
%To eliminate the auxiliary SWAP gates in $U_e$, the qubit map to the quantum computer is shown in Fig.(\ref{fig_ibmq}a), where there is no physical connection between qubits $q_{1,2,3,4}$.
%On the other hand, in the operation estimating $U_{dis}$ and $exp(i\lambda V)$, there are dozens of CNOT gates applied on qubits $q_{1,2,3,4}$.
%Therefore, if we insist on running the whole proposed circuit on a quantum computer, quantities of auxiliary SWAP gates would be required, which consequently could lead to considerable error in the output.
%Alternatively, here we attempt to estimate the corrections with a hybrid calculation.
%The contributions of $U_e$ are estimated on the quantum computer separately, while the operations $U_{dis}$ and $exp(i\lambda V)$ are estimated via the simulator classically.

%The hybrid results are much great than the simulation results (marked with purple triangles in Fig.(\ref{fig_simu}c,d)).

\section{Discussion}
\label{Discussion}
In conclusion, we propose a general quantum circuit estimating both the energy and eigenstates corrections with perturbation theory (PT).
The quantum approach is demonstrated with application to the two-site extended Hubbard model, where we present numerical simulations based on Qiskit.
Furthermore, we implement the proposed circuit on the IBM 27-qubit quantum computer, `ibmq\_montreal', demonstrating the practicality of estimating PT corrections with quantum hardware.
Compared to classical PT, the quantum method is always more efficient in estimating the second-order energy correction $E_{n}^{(2)}$ for complex systems.
When studying complex systems with considerable degeneracy, the quantum method is also more efficient in estimating the first-order eigenstate correction $|\psi_n^{(1)}\rangle$.
Moreover, all parameters in the quantum circuit are determined directly by the given Hamiltonian, eliminating any training or optimization process.
Our work provides a new approach to studying complex systems with quantum devices, making it possible to implement PT-based methods on with a quantum computer on a wide variety of problems in chemistry and physics.

\section{Materials and Methods}
\label{Materials and Methods}
\subsection{Time complexity}
\label{Time complexity}
In this section, we will briefly analyze the time complexity of our method and compare it with classical PT.
The unperturbed energy and eigenstates are always required in PT methods.
When the unperturbed Hamiltonian $H_0$ is not available or hard to compute, the popular quantum variational circuit would be a better choice.
Here we assume that the unperturbed Hamiltonian is already well-studied so that $E^{(0)}_{n}$, $|\psi_n^{(0)}\rangle$ are given initially, and the time complexity to derive $E^{(0)}_{n}$, $|\psi_n^{(0)}\rangle$ is not included in the following discussion.

Consider a system with $2^N$ basis states and $L$ different energy levels, where $L\leq 2^N$.
Due to the existence of degeneracy, $L$ can be sometimes much less than the number of basis states.
One example can be found in Fig.(\ref{fig_ue}A), where there are 16 basis states but only 6 different energy levels.
Referring to the quantum circuit shown in Fig.(\ref{fig_main}), we need $N$ qubits representing the system with $2^N$ basis states.
In this paper, we studied the extended Hubbard model, which contains on-site energy and interactions between nearest neighbors, leading to $\mathcal{O}(N)$ time complexity simulating the perturbation $V$ or $exp(i\lambda V)$.
As for more complicated systems with long-range interactions, theoretically no more than $\mathcal{O}(N^2)$ would be required to simulate the perturbation.
{Assuming interactions between nearest and next-nearest neighbor sites in an N-site model, the number of pairs of sites is $N(N-1)/2$. Simulating each interaction would require multiple two-qubit gates, resulting in an overall time complexity of no more than $\mathcal{O}{(N^2)}$ for simulating these long-range interactions.}
% Consider an N-site model where there are interactions between both nearest and next-nearest neighbor sites: there are in total $N(N-1)/2$ pairs of sites.
% Each interaction could be simulated with several two-qubit gates.
% Therefore, the overall time complexity simulating such long-range interactions is no more than $\mathcal{O}{(N^2)}$.

In the study of the extended Hubbard model, we construct the operator $U_{dis}$ with quantum Fourier transform on the nearest neighbors and two special multi-controlled rotation gates.
The Fourier transform part requires $\mathcal{O}(N)$ time complexity, while the multi-controlled rotation gates with $N$ controlled qubits can be decomposed into $\mathcal{O}(N^2)$ CNOT gates and single qubit rotation gates\cite{barenco1995elementary}, leading to $\mathcal{O}(N^2)$ time complexity.
Including all of the above, the time complexity to estimate the perturbation terms $\langle \psi_m^{(0)}|V|\psi_n^{(0)}\rangle$ is no more than $\mathcal{O}(N^2)$.
To estimate a quantum output within error $\epsilon$, $\mathcal{O}(\frac{1}{\epsilon^2})$ measurement time is required\cite{gilyen2019quantum}.
In total, the time complexity estimating the first-order energy correction $E_{n}^{(1)}$ is $\mathcal{O}(N^2/\epsilon^2)$.
Meanwhile, there are $L$ multi-controlled rotation gates in $U_e$, leading to $\mathcal{O}(LN^2)$ time complexity.
Therefore, the time complexity estimating the first-order eigenstate correction and second-order energy correction is $\mathcal{O}(LN^2/\epsilon^2)$.

In contrast, classical PT estimates the corrections as shown in Eq.(\ref{eq_energy1}, \ref{eq_state1}, \ref{eq_energy2}).
When estimating the first-order energy correction $E_{n}^{(1)}$, only 1 term is calculated.
However, $\mathcal{O}(2^N)$ terms are calculated to estimate the first order eigenstate correction $|\psi_n^{(1)}\rangle$, and a further $\mathcal{O}(4^N)$ terms are calculated to estimate the second order energy correction $E_{n}^{(2)}$.
The number of basis states dominates the time complexity of classical PT methods.
Compared with classical PT, our quantum version does not show speed up when estimating the first-order energy correction $E_{n}^{(1)}$.
However, our quantum circuit can also generate the quantum state of the first-order eigenstate correction $|\psi_n^{(1)}\rangle$.
When studying a complex system with considerable degeneracy, we have $L\ll 2^N$, and the quantum methods can lead to speedup when estimating $|\psi_n^{(1)}\rangle$.
Most importantly, the quantum version leads to speedup when estimating the second order energy correction $E_{n}^{(2)}$ of complex systems with large size, since for large $N$ values, we have $\mathcal{O}(LN^2/\epsilon^2)<\mathcal{O}(4^N)$.

\subsection{Applications}
\label{Applications}
In Sec. (\ref{Applications on Extended Hubbard Model}) and Sec.(\ref{Implementation on Quantum Computer}), the proposed quantum circuit design and implementation on real quantum hardware are demonstrated in detail, with application to the extended 2-site Hubbard model.
In this section, we would like to expand on the class of problems to which our method could be applied.

In addition to the simple 2-site Hubbard model, our proposed method is applicable to other strongly correlated quantum systems.
There are three key operations in our proposed quantum circuit, $U_{dis}$, $exp(i\lambda V)$, and $U_e$.
In Sec.(\ref{Applications on Extended Hubbard Model}), we present a universal design of $U_e$ with multi-controller gates.
Similarly, given known perturbation $V$, we could design $exp(i\lambda V)$ with Trotter decomposition.
Meanwhile, Verstraete et. al developed the explicit quantum circuits that diagonalize the dynamics of strongly correlated quantum systems with a Bogoliubov transformation and quantum Fourier transformation\cite{verstraete2009quantum}, with which $U_{dis}$ could be generalized to these quantum systems.
Our proposed quantum circuit therefore could be applied to other strongly correlated quantum systems.
As an example, in the Supplementary Materials, we present another application, to a Heisenberg XY chain.

Furthermore, as PT is always a powerful tool for chemists solving many quantum chemistry problems, our proposed quantum circuit could also be applied to electronic structure calculations for atoms and molecules.
For instance, M\o ller–Plesset perturbation theory (MPPT)\cite{moller1934note} is a typical post–Hartree–Fock ab initio method in the field of computational chemistry, where a Hartree–Fock (HF) calculation is used as the starting point, and the difference between the exact Hamiltonian and the HF one is included as a perturbation.
In recent years, we have witnessed a multiplicity of quantum theoretical and experimental tools for the prediction of molecular properties and chemical reactions pathways and structure, especially with the HF method.
In 2020, Google AI Quantum successfully obtained the Hartree-Fock wave function for a linear chain of 12 hydrogen atoms with a variational quantum eigensolver (VQE) simulation on their Sycamore quantum processor\cite{google2020hartree}.
Such advances bring us more promising applications, making it possible to develop quantum circuits for MPPT calculations,
where the HF results could be obtained from quantum devices with VQE simulation, and the PT calculations from our proposed quantum circuit.
%In a nutshell, the proposed quantum circuit could be applied to various correlated quantum systems. 
In summary, the proposed general quantum circuit could be applied to various strongly correlated many-body quantum systems.

\section*{Acknowledgements}
\label{Acknowledgements}
The authors would like to acknowledge Dr. Manas Sajjan, Dr. Kale, Sumit Suresh, Dr. Rishabh Gupta and Dr. Bibek Pokharel for fruitful discussions.
\newline
{\bf Funding:}
We acknowledge funding by the U.S. Department of Energy (Office of Basic Energy Sciences) under Award No. DE-SC0019215, and the  National Science Foundation under Award No. 1955907.
This work was also  supported by the U.S. Department of Energy (DOE),
Office of Science through the Quantum Science Center (QSC),
a National Quantum Information Science Research Center.
\newline
{\bf Author contributions:}
S.K. and J.L. designed the model and the computational framework.
J.L. carried out the implementation, and performed the numerical simulations and experiments.
All authors discussed the results and wrote the paper. 
S.K. was in charge of the overall direction and planning.
\newline
{\bf Competing interests:}
The authors declare that they have no competing interests.
\newline
{\bf Data and materials availability:}
All data needed to evaluate the conclusions in the paper are present in the paper and/or the Supplementary Materials.

\bibliography{ref}

%Supplementary Materials
\clearpage
\newcommand{\beginsupplement}{
    \setcounter{section}{0}
    \renewcommand{\thesection}{S\arabic{section}}
    \setcounter{equation}{0}
    \renewcommand{\theequation}{S\arabic{equation}}
    \setcounter{figure}{0}
    \renewcommand{\thefigure}{S\arabic{figure}}}
\beginsupplement

\section*{Supplementary Materials}
\subsection*{
A. An alternative implementation of the main circuit
}

\begin{figure}[ht]
    \centering
    \includegraphics[width=0.85\textwidth]{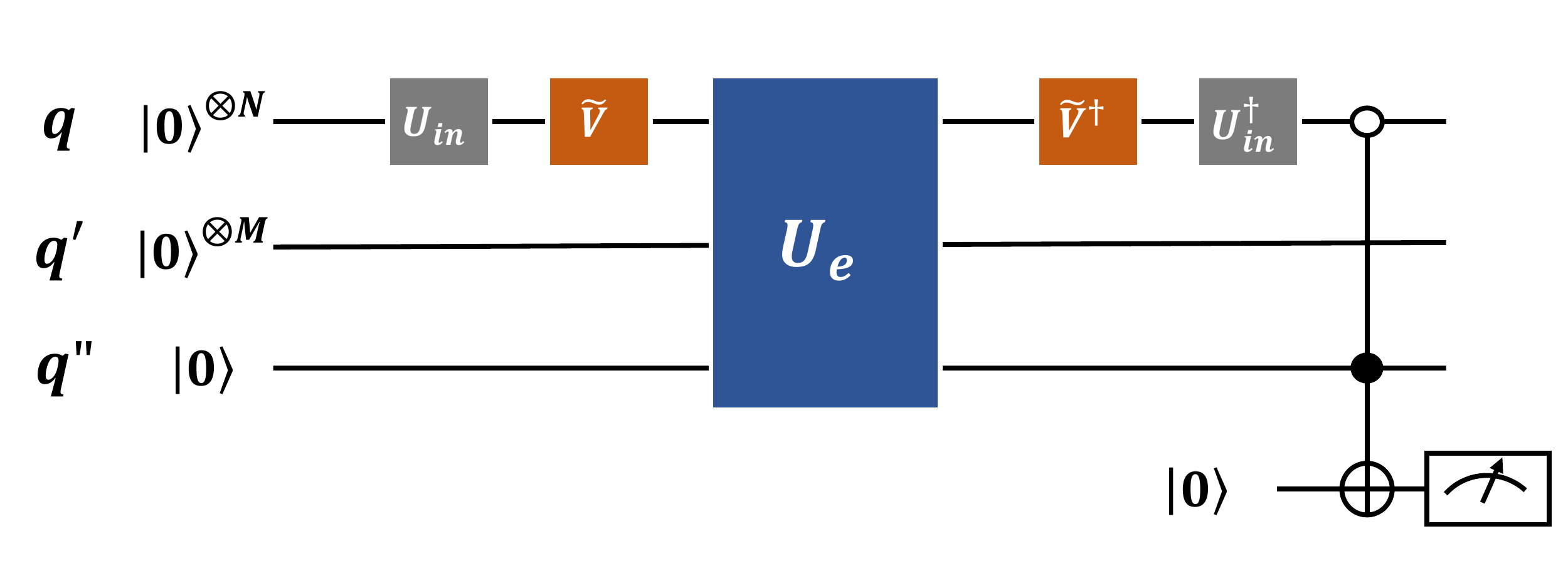}
    \caption{
    {\bf Scheme of the alternative implementation of the main circuit.}
    }
    \label{figs_main}
\end{figure}

In the main article, the Repeat-Until-Success(RUS) strategy is performed to obtain the first order eigenstate correction $|\psi^{(1)}_n\rangle$, which is then used for the next circuit estimating $E_n^{(2)}$.
The RUS process includes the measurement on intermediate states, which raises additional requirements for the experimental apparatus.
Sometimes it might be difficult to apply successive operators after intermediate measurements.
In this case, we can end at state $|\phi_{III}\rangle$ shown in Fig.(1b), with which the first order eigenstate correction $|\psi^{(1)}_n\rangle$ could be estimated.
In Fig.(\ref{figs_main}), we present an optional circuit implementation to estimate the second order energy correction $E_n^{(2)}$ without the RUS process.
Differing from the original circuit shown in Fig.(1b), $q''$ is not measured after $U_e$.
Instead, $q''$ performs as control qubit in the multi-controlled NOT  gate at the end.
In this way, we could estimate $E_n^{(2)}$ without measurements on the intermediate states.

\subsection*{
B. Application to a Heisenberg XY chain
}

Here we give another application of our quantum circuit for perturbation theory (PT) methods.
Consider the Heisenberg XY chain, whose Hamiltonian is described as
\begin{equation}
    H = \sum_{j}\sigma_{j}^x\sigma_{j+1}^x + \gamma\sum_j\sigma^z_{j}
    +\lambda\sum_{j}\sigma_{j}^y\sigma_{j+1}^y
\end{equation}
where $\gamma\sum_j\sigma^z_{j}$ represents an external transverse magnetic field.
We assume that the $YY$ interaction is much weaker {as compared} to the $XX$ interaction, so that $\lambda\ll1$, and the $YY$ interaction terms can be regarded as perturbations.

\begin{figure}[ht]
    \centering
    \includegraphics[width=0.85\textwidth]{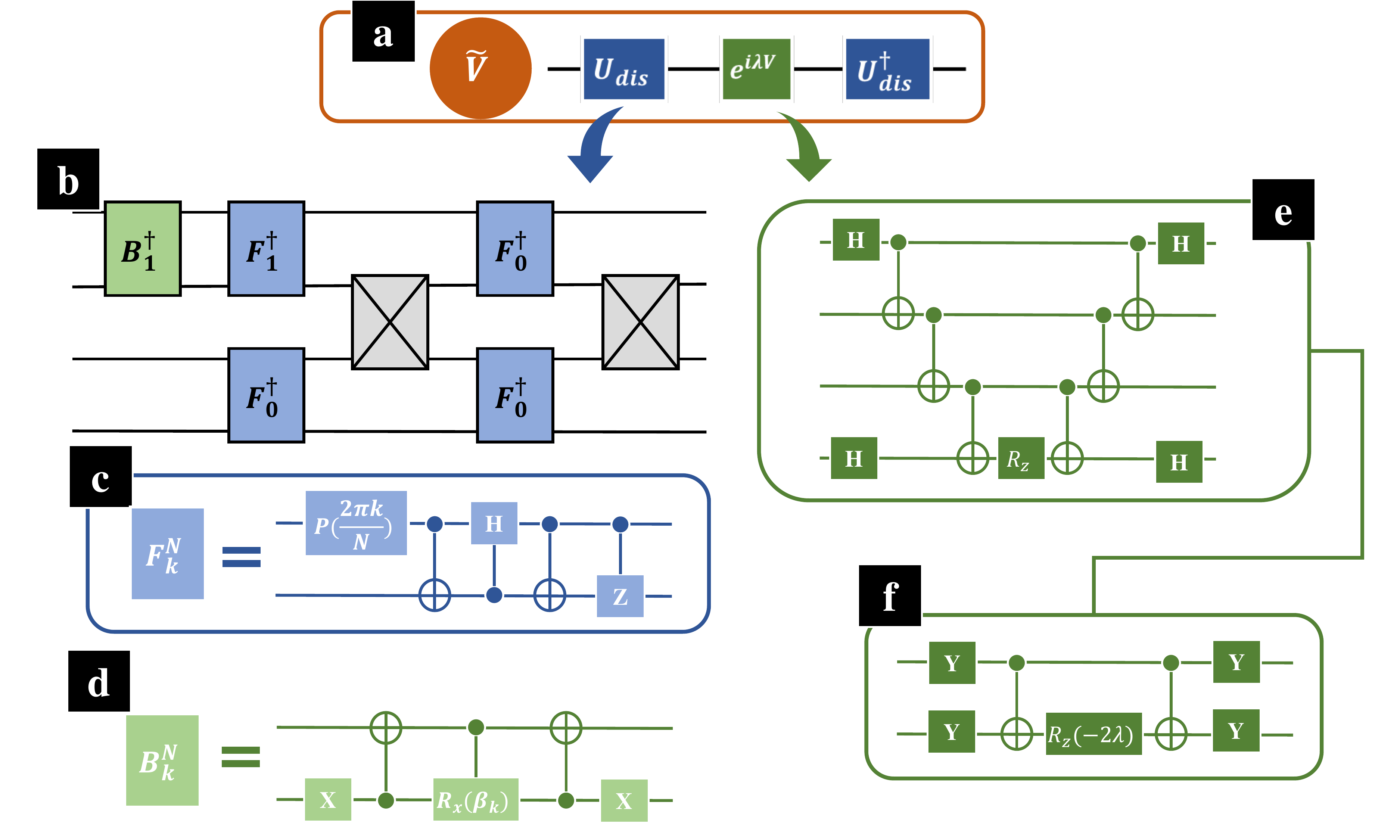}
    \caption{
    {\bf Scheme of the quantum circuit implementation for the Heisenberg XY chain.}
    \\
    (a.) The decomposition of $\Tilde{V}$.
    As we are studying a new system, the operator $U_{dis}$ and $\exp{(i\lambda V)}$ are different than eg Fig.(\ref{fig_main}D,H).
    (b.) Structure of $U_{dis}$ that converts the computational basis $|n\rangle$ into the unperturbed eigenstates $|\psi_n^{(0)}\rangle$.
    Operator $B$ represents a Bogoliubov transformation, while $F$ represents the quantum Fourier transform.
    The grey box is a fermionic swap gate.
    (c.) Structure of the quantum Fourier transform.
    (d.) Structure of the Bogoliubov transformation.
    With Trotter decomposition, operator $\exp{(i\lambda V)}$ could be decomposed into the $ZZ$ interaction shown in (e) and the interaction at boundary as shown in (f).
    }
    \label{figs_ising}
\end{figure}

The decomposition of $\Tilde{V}$ can be found in Fig.(\ref{figs_ising}).
As we are studying a new system, the operator $U_{dis}$ and $\exp{(i\lambda V)}$ are different than the previous Hamiltonian.
The structure of $U_{dis}$ is presented in Fig.(\ref{figs_ising}b), which converts the computational basis $|n\rangle$ into the unperturbed eigenstates $|\psi_n^{(0)}\rangle$, leading to $U_{dis}|n\rangle=|\psi_n^{(0)}\rangle$.
Operator $B$ represents a Bogoliubov transformation, while $F$ represents the quantum Fourier transform.
The grey box is a fermionic swap gate.
The structure of the quantum Fourier transform is shown in Fig.(\ref{figs_ising}c),
while the structure of the Bogoliubov transformation can be found in Fig.(\ref{figs_ising}d).
With Trotter decomposition, operator $\exp{(i\lambda V)}$ can be decomposed into the $ZZ$ interaction shown in Fig.(\ref{figs_ising}e) and the interaction at the boundary as shown in Fig.(\ref{figs_ising}f).
More details about the operator $U_{dis}$ for Ising type Hamiltonians can be found in \cite{verstraete2009quantum, cervera2018exact}.

Once the unperturbed energy levels $E_n^{(0)}$ are given, $U_e$ can be constructed following the method discussed in Sec.(2) of the main article.
For simplicity, we will not repeat the whole process.

\subsection*{
C. More discussion about $U_e$
}
\begin{figure}[ht]
    \centering
    \includegraphics[width=0.85\textwidth]{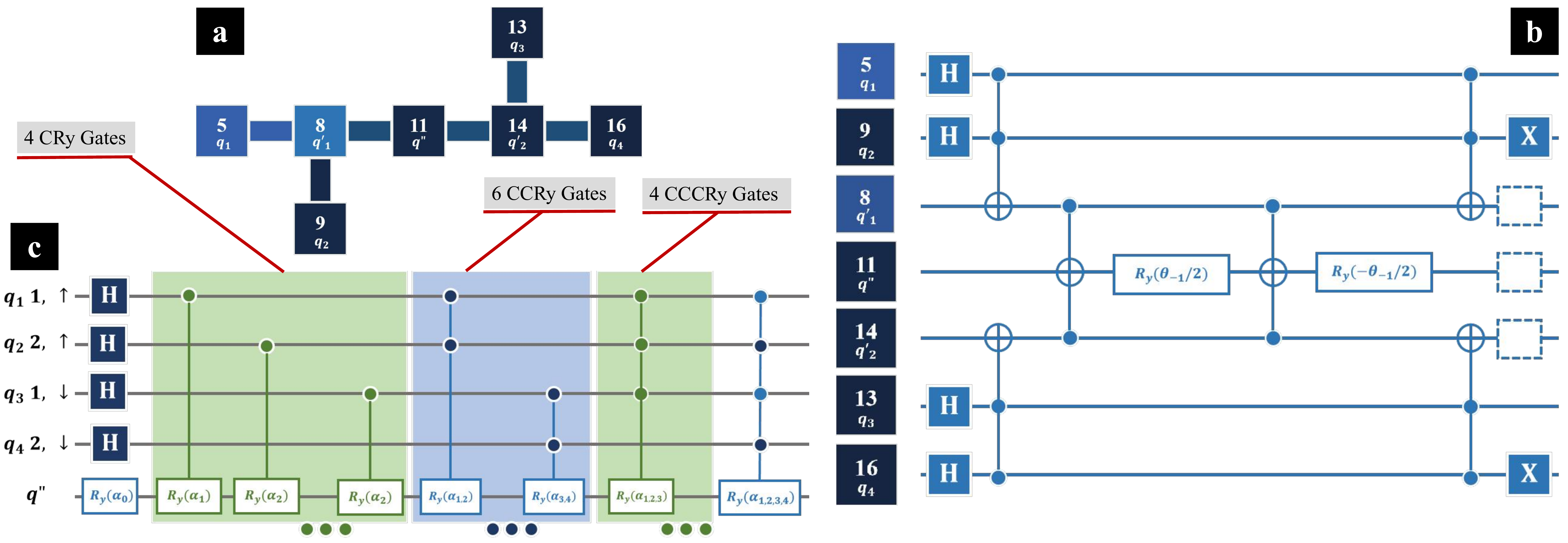}
    \caption{
    {\bf Details of $U_e$.}
    (a) The mapping on quantum computer 'ibmq\_montreal', each square indicating a qubit on the quantum computer.
    (b) Decomposition of the multicontroller gate with 4 control qubits, along with the Hadamard gates on $q_{1,2,3,4}$, and NOT gates on $q_{2,4}$, corresponding to the gate with green background in Fig.(4c) (Step 3).
    (c) An equivalent decomposition of $U_e$, where there are 1 Ry gate, 4 CRy gates, 6 CCRy gates, 4 CCCRy gates and 1 CCCCRy gate (a CRy gate contains one controller qubit, a CCRy gate contains two and so on).
    The Hadamard gates on $q_{1,2,3,4}$ are included to test the performance.
    }
    \label{figs_ue}
\end{figure}

%In this section, we would like to present an alternative construction of $U_e$ with quantum variational circuit.
%As shown in Fig.(\ref{figs_ue}) we present an alternative approach construction of $U_e$, which is implemented by quantum variational circuit.
%Differing from the popular quantum variational circuits, $q$ qubits, the input of $U_e$, should be kept unchanged.
%Thus, $M$ ancilla qubits $q'$ are included.
%As shown in the dashed box of Fig.(\ref{figs_ue}), there are controlled rotation Y gates between each $q$ qubit (as control qubit) and each $q'$ qubit (as target).
%For simplicity, the controlled rotation Y gates with the same control qubit are plotted in a branch. 
%Then there are controlled rotation Y gates between $q'$ and $q''$, with $q'$ qubits as control qubit and $q''$ itself as target.
%$\theta$ of these control rotation Y gates are variational.
%Similarly, the ${\vartheta}$ are variational.
%Finally, the inverse operator of ones in the dashed box should be applied.
%More variational layers could be added on $q'$, pursuing more accurate results.
%The variational version would be able to reduce the scale and the depth of $U_e$ in some cases, even though much time is required fulfilling the training and optimizing process.
%Therefore, the variational version of $U_e$ is not applied in the main article, but left here for those with interest.
In this section, we would like to present more details about $U_e$.
As shown in Fig.(4c), $U_e$ could be decomposed into a few multicontroller gates along with several simple single qubit gates.
When implementing the multicontroller gates on a quantum device, we apply the decomposition as shown in Fig.(\ref{figs_ue}b), which corresponds to the multicontroller gate with green background as shown in Fig.(4c).
$q'_{1,2}$ are introduced to connect control qubits $q_{1,2,3,4}$ with the target $q"$.
There are CCNOT gates (or Toffoli gates) applied among $q_{1},q_2$ and $q'_1$, similarly $q_{3},q_4$ and $q'_2$.
Then a CCRy gate is applied among $q'_{1,2}$ and $q"$, which is decomposed into two CCNOT gates and two Ry gates.
Finally, there are two CCNOT gates applied among qubits $q_{1,2,3,4}$ and $q'_{1,2}$, ensuring the later ones are reversed to their initial states.
In addition to the decomposition of the multicontroller gate, there are some simple Hadamard gates or NOT gates in Fig.(\ref{figs_ue}b), which correspond to the same ones in Fig.(4c).
The CCNOT gates pairs as shown in Fig.(4c) could be decomposed into the circuit shown in Fig.(4d), where the operations with grey background cancel out.
Therefore, in the implementation on real quantum devices as shown in Fig.(\ref{figs_ue}a), there are no more quantum SWAP gates required.
In the experiment on real quantum devices, we notice that the CCNOT gates mainly cause error on the target qubit.
For instance, when we test the circuit as shown in Fig.(\ref{figs_ue}b), all qubits are initialized at ground state $|0\rangle$.
Theoretically qubits $q'_{1,2}$ should always be $|0\rangle$ at the end, yet in experiment they are often not.
Additional Ry gates can then be added on $q'_{1,2}$, and similarly on $q"$, as in the dashed boxes shown in  Fig.(\ref{figs_ue}b).

In Fig.(2b) and Fig.(4c), the $U_e$ are decomposed into multicontroller gates along with several simple single qubit gates.
Due to the degeneracy in our example, only 7 multicontroller gates are required.
Here we present an equivalent decomposition of $U_e$ as shown in Fig.(\ref{figs_ue}), where there are 1 Ry gate, 4 CRy gates, 6 CCRy gates, 4 CCCRy gates and 1 CCCCRy gate (a CRy gate contains one controller qubit, a CCRy gate contains two and so on).
The design in Fig.(\ref{figs_ue}) is also universal, but leads to less time complexity when there is no degeneracy.
In Tab.(\ref{tabs_result}), we present the probabilities to get various results.
Consider the quantum circuit shown in Fig.(\ref{figs_ue}c), and measure $q_{1,2,3,4}$ and $q"$.
The first four columns indicate the measurement results of $q_{1,2,3,4}$, and the last column represents the probability to obtain the corresponding results on $q_{1,2,3,4}$ while obtaining result 1 on $q"$.

\begin{table}[h]
    \centering
    \begin{tabular}
{ 
|c|c|c|c|c|}
  %| >{\centering\arraybackslash}X 
  %| >{\centering\arraybackslash}X 
  %| >{\centering\arraybackslash}X 
  %| >{\centering\arraybackslash}X 
  %| >{\centering\arraybackslash}X | }
 \hline
 $q_1$ &$q_2$ &$q_3$ &$q_4$ &Probability to get this result \\
 \hline
 0  &0  &0 &0 &$\frac{1}{16}\sin^2{\left(\frac{\alpha_0}{2}\right)}$ \\
\hline
0  &0  &0 &1 &$\frac{1}{16}\sin^2{\left(\frac{\alpha_0+\alpha_4}{2}\right)}$ \\
\hline
0  &0  &1 &0 &$\frac{1}{16}\sin^2{\left(\frac{\alpha_0+\alpha_3}{2}\right)}$ \\
\hline
0  &0  &1 &1 &$\frac{1}{16}\sin^2{\left(\frac{\alpha_0+\alpha_3+\alpha_4+\alpha_{3,4}}{2}\right)}$ \\
\hline
0  &1  &0 &0 &$\frac{1}{16}\sin^2{\left(\frac{\alpha_0+\alpha_2}{2}\right)}$ \\
\hline
0  &1  &0 &1 &$\frac{1}{16}\sin^2{\left(\frac{\alpha_0+\alpha_2+\alpha_4+\alpha_{2,4}}{2}\right)}$ \\
\hline
0  &1  &1 &0 &$\frac{1}{16}\sin^2{\left(\frac{\alpha_0+\alpha_2+\alpha_3+\alpha_{2,3}}{2}\right)}$ \\
\hline
0  &1  &1 &1 &$\frac{1}{16}\sin^2{\left(\frac{\alpha_0+\alpha_2+\alpha_3+\alpha_4+\alpha_{2,3}+\alpha_{2,4}+\alpha_{3,4}+\alpha_{2,3,4}}{2}\right)}$ \\
\hline
1  &0  &0 &0 &$\frac{1}{16}\sin^2{\left(\frac{\alpha_0+\alpha_1}{2}\right)}$ \\
\hline
1  &0  &0 &1 &$\frac{1}{16}\sin^2{\left(\frac{\alpha_0+\alpha_{1}+\alpha_{4}+\alpha_{1,4}}{2}\right)}$ \\
\hline
1  &0  &1 &0 &$\frac{1}{16}\sin^2{\left(\frac{\alpha_0+\alpha_{1}+\alpha_{3}+\alpha_{1,3}}{2}\right)}$ \\
\hline
1  &0  &1 &1 &$\frac{1}{16}\sin^2{\left(\frac{\alpha_0+\alpha_{1}+\alpha_{3}+\alpha_{4}+\alpha_{1,3}+\alpha_{1,4}+\alpha_{3,4}+\alpha_{1,3,4}}{2}\right)}$ \\
\hline
1  &1  &0 &0 &$\frac{1}{16}\sin^2{\left(\frac{\alpha_0+\alpha_{1}+\alpha_{2}+\alpha_{1,2}}{2}\right)}$ \\
\hline
1  &1  &0 &1 &$\frac{1}{16}\sin^2{\left(\frac{\alpha_0+\alpha_{1}+\alpha_{2}+\alpha_{4}+\alpha_{1,2}+\alpha_{1,4}+\alpha_{2,4}+\alpha_{1,2,4}}{2}\right)}$ \\
\hline
1  &1  &1 &0 &$\frac{1}{16}\sin^2{\left(\frac{\alpha_0+\alpha_{1}+\alpha_{2}+\alpha_{3}+\alpha_{1,2}+\alpha_{1,3}+\alpha_{2,3}+\alpha_{1,2,3}}{2}\right)}$ \\
\hline
1  &1  &1 &1 &$\frac{1}{16}\sin^2{\left(\frac{\alpha_0+\alpha_{1,2,3,4}}{2}+\frac{1}{2}\sum_{a}{\alpha_a}+\frac{1}{2}\sum_{a,b}{\alpha_{a,b}+\frac{1}{2}\sum_{a,b,c}{\alpha_{a,b,c}}}\right)}$ \\
\hline
\end{tabular}
    \caption{{\bf Probabilities to get various results.}
    Consider the quantum circuit shown in Fig.(\ref{figs_ue}c), and measure $q_{1,2,3,4}$ and $q"$.
    The first four columns indicate the measurement results of $q_{1,2,3,4}$, and the last column represents the probability to obtain the corresponding results on $q_{1,2,3,4}$ while obtaining result 1 on $q"$.
    }
    \label{tabs_result}
\end{table}
Additionally, due to the existence of multiple sources of noise, we can further reduce the number of multicontroller gates.
Some multicontroller Ry gates with small parameters are extremely sensitive, and their contribution is even less than their average error.
In our experiment, multicontroller gates with $\theta_{gs}$ and ${\theta_1}$ lead to more error than contribution in the overall result, and we can get a better result after excluding these gates, as shown in Fig.(3c,d).

Another construction of $U_e$ might be useful in certain cases, where the terms are approximated with Fourier series,
\begin{equation}
    \frac{C}{E_0-E_n} = \sum_{m=1}a_m\cos(mn + b_m)
    \label{eqs_fourier}
\end{equation}
The parameters $\{a_m, b_m\}$ guarantee that Eq.(\ref{eqs_fourier}) works for all possible $n$ (when $n=0$ the left part is set to 0). 
In our recent work\cite{li2021universal} we present a quantum circuit estimating the Fourier series as shown in Eq.(\ref{eqs_fourier}), which would be helpful especially when $\frac{C}{E_0-E_n}$ is periodic with degeneracies.
\\\\
%Detailed circuit

\subsection*{
D. Detailed quantum circuit
}

\begin{figure}[ht]
    \centering
    \includegraphics[width=0.85\textwidth]{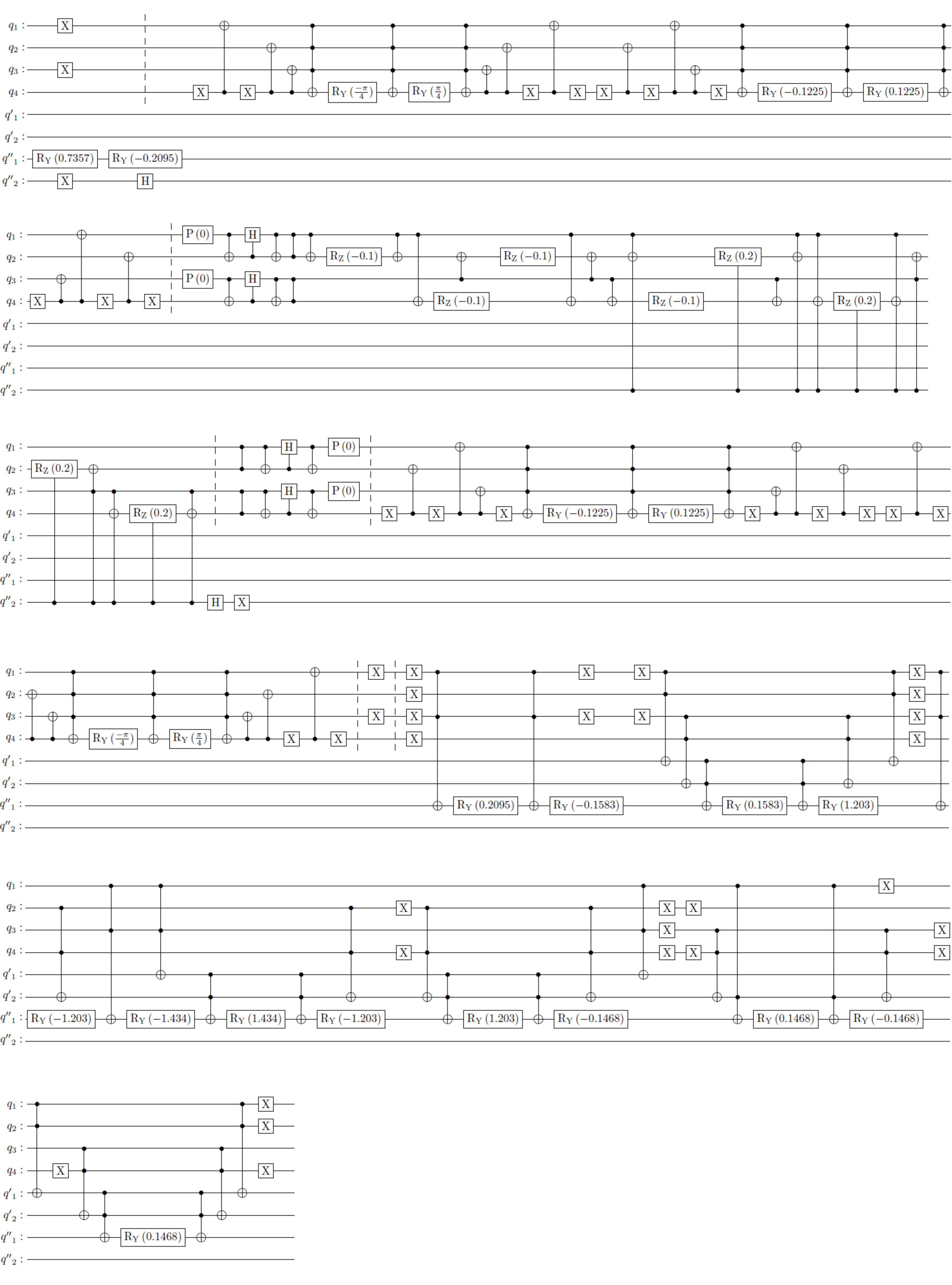}
    \caption{
    {\bf Scheme of the full quantum circuit estimating the first order eigenstate correction.}
    The first 4 qubits $q_1, q_2, q_3, q_4$ represent the system we are studying.
    $q'_1, q'_2$ are included to construct multi controlled rotation gates in $U_e$.
    $q"_1$ corresponds to the $q"$ shown in Fig.(1b).
    $q"_2$ is the ancilla qubit included in the improved circuit estimating perturbation, as shown in Fig.(3c).
    Here we set $\lambda = 0.1$.
    In total, there are more than 80 single qubit gates, around 50 two qubit gates (CNOT, CRy, and SWAP gates), 38 Toffoli gates (also CCNOT gates) and 12 CCCNOT gates.
    }
    \label{figs_full_circuit}
\end{figure}

In Fig.(\ref{figs_full_circuit}) we present the detailed quantum circuit estimating the first order eigenstate correction $\psi_0^{(1)}$ for the ground state.
The first 4 qubits $q_1, q_2, q_3, q_4$ represent the system we are studying.
$q'_1, q'_2$ are included to construct multicontrolled rotation gates in $U_e$.
$q"_1$ corresponds to the $q"$ shown in Fig.(1b).
$q"_2$ is the ancilla qubit included in the improved circuit estimating perturbation, as shown in Fig.(3c).
After the whole operation, $q"_1$ is measured.
If result $|1\rangle$ is obtained, the first order eigenstate correction could be estimated as discussed in Sec.(1) in the main article.
In Fig.(\ref{figs_full_circuit}), we set $\lambda = 0.1$.
In total, there are more than 80 single qubit gates, around 50 two qubit gates (CNOT, CRy, and SWAP gates), 38 Toffoli gates (also CCNOT gates) and 12 CCCNOT gates.
All the key operations $U_{dis}$, $U_e$, and ${\exp(i\lambda V/2)-\exp{(-i\lambda V/2)}}$ are included in the quantum circuit estimating the first order eigenstate correction.

%\bibliographystylelatex{plain}
%\bibliographylatex{refs}
\end{document}